\documentclass[referee]{aa}
\usepackage{graphicx}
\begin{document}
\title{Magnetohydrodynamic Model of Equatorial Plasma Torus
\\ in Planetary Nebulae}
\author{K.H. Tsui\thanks{\email{tsui$@$if.uff.br}}}
\institute{Instituto de F\'{i}sica - Universidade Federal Fluminense
\\Campus da Praia Vermelha, Av. General Milton Tavares de Souza s/n
\\Gragoat\'{a}, 24.210-346, Niter\'{o}i, Rio de Janeiro, Brasil.}
\date{Received / Accepted}
\baselineskip 18pt

\abstract
{}{Some basic structures in planetary nebulae are modeled as
 self-organized magnetohydrodynamic (MHD) plasma configurations
 with radial flow.}
 {These configurations are described by time self-similar dynamics,
 where space and time dependences of each physical variable
 are in separable form. Axisymmetric toroidal MHD plasma configuration
 is solved under the gravitational field of a central star of mass $M$.}
 {With an azimuthal magnetic field, this self-similar MHD model
 provides an equatorial structure in the form of an axisymmetric
 torus with nested and closed toroidal magnetic field lines.
 In the absence of an azimuthal magnetic field, this formulation
 models the basic features of bipolar planetary nebulae.
 The evolution function, which accounts for the time evolution
 of the system, has a bounded and an unbounded evolution track
 governed respectively by a negative and positive energy density
 constant $H$.}{}

\keywords{planetary nebulae : equatorial density enhancement, 
 planetary nebulae : bipolar nebulae}
\titlerunning{Equatorial Plasma Torus}
\authorrunning{Tsui}
\maketitle

\newpage
\section{Introduction}

Planetary nebulae are believed to be intermediate-mass stars
 that are expelling their outer layers, at slow velocity,
 during the asymptotic giant branch (AGB) phase.
 As the temperature of the star increases, the ejected material
 becomes tenuous but moves with increasing speed.
 The hydrodynamic interaction of these slow and fast winds,
 at the intersection of their trajectories, generates shock waves
 that create the characteristic morphologies of planetary nelulae
 (Dyson and de Vries 1972, Kwok et al 1978).
 The outward-shock compresses and heats the slow dense wind
 that is ahead,
 and the inward-shock decelerates and heats the fast wind behind.
 This interacting-wind model reproduces the spherical features
 of planetary nebulae.
 In contrast, the elliptical and bipolar features may be created
 by a dense toroidal cloud in the equatorial plane
 (Kahn and West 1985, Mellema et al 1991).
 Although toroidal clouds are imaged by high-resolution
 instruments, how such structures formed during the AGB phase
 is still a debated issue.
 Besides spherical, elliptical, and bipolar features,
 high-resolution images have revealed point-symmetric features
 (Miranda and Solf 1992, Lopez et al 1993, Balick et al 1993).
 These point-symmetric features, plus the detection of significant
 magnetic fields inside central stars (Jordan et al 2006),
 appear to call for a magnetohydrodynamic (MHD) approach
 to the study of planetary nebulae
 (Pascoli 1993, Chevalier and Luo 1994, Garcia-Segura 1997,
 Bogovalov and Tsinganos 1999, Matt et al 2000, Gardiner and Frank 2001).
 In this paper, we present such an MHD analysis,
 which is able to recreate axisymmetric structures
 such as an equatorial plasma torus and a bipolar nebula.
 We consider an MHD plasma that simulates a radial-flow explosion
 in spherical coordinates, and we solve for solutions that are
 self-similar in time (Sommerfeld 1950, Landau and Lifshitz 1978).

Self-similar MHD analysis was pioneered by Low (1982a,b, 1984a,b),
 in astrophysical stellar-envelope applications.
 The association of self-similar solutions to eruptive phenomena
 can be best demonstrated by the generation of self-similar shocks
 in powerful atmospheric detonations. The shocks are the results
 of self-organization, through dissipative processes, following chaos.
 The shock is built up as the gas expands outward, and
 it takes its fully developed form beyond a certain radius.
 These ordered structures could have a scale invariance
 that lowers the dimensionality of the time-dependent fluid system,
 and could exist between one spatial coordinate and time.
 This invariance is identified as the self-similar parameter,
 and similarity is temporal.
 Time evolution would be a separable factor, and, as a result,
 the time and space dependencies of self-similar
 variables are represented in separable form.
 For steady-state systems, the scale invariance could be between
 one spatial coordinate and another,
 and the similarity is said to be spatial.

Self-similarity is often regarded as a method to derive
 specific types of time-dependent or steady-state solutions.
 In this paper, we consider self-similarity to be a property
 of self-organized states,
 at least some of which are derived from turbulence.
 It is essential to note that the appearance of self-organized
 states, in fluids and magnetofluids, owes itself to the existence
 of multiple quadratic invariants in the absence of dissipations
 (Hasegawa 1985, Biskamp 1993).
 In the case of MHD fluids, the invariants are the total energy
 density (magnetic, plasma, and thermal), magnetic helicity,
 and in the case of incompressible fluids, cross-helicity.
 In the presence of dissipation, these invariants could undergo
 constrained changes, for example, from an MHD configuration,
 permitted by the MHD equations, to another topologically accessible
 configuration of lower energy.
 It is important to point out
 that these configurations are isolated in ideal cases.
 The continuous topological transformation from one to
 another configuration is only possible through dissipative paths
 such as magnetic reconnections.
 Mathematically speaking, this evolution amounts
 to the application of the variational principle,
 on the total energy density, under the constraint of
 global magnetic helicity conservation (Hasegawa 1985).
 The exact processes of dissipation,
 in taking the system to self-organization,
 need not be specified under the variational description.
 For this reason, MHD systems have the tendency to develop
 self-organized states with structural stability.

Since the self-similar method solves for the end configurations,
 directly from the governing equations,
 there is no need to specify the initial conditions.
 The plasma would find its way through magnetic reconnection,
 and other dissipative means,
 to reach one of the possible end configurations.
 The only condition is that the initial configuration
 is dissipatively transformable to the final configuration.
 If the initial conditions are close to those of the end
 configuration, the time to reach the end configuration is brief.
 Naturally, there are many end configurations, and only one of them
 will be selected by dissipative reorganization. For this reason,
 we cannot predict which self-similar state the system will reach.
 We can only match a system to a given self-similar solution
 that bears some resemblance.

One of the fundamental obstacles to visualizing self-similar
 solutions is an attempt to associate a given initial
 configuration with a specific self-similar end solution.
 It is possible to predict the final state of ideal MHD
 systems, in particular, by iterating equations of motion
 for an initial configuration.
 In the presence of dissipations, it can, however, be difficult
 to predict the final state, because of intermediate, chaotic
 events that occur while states are being reorganized.
 The initial configuration could undergo topological changes,
 through dissipative processes, such as magnetic reconnections,
 subject to the constraint of quadratic invariants.
 To describe the self-organized states arising from chaos,
 we have to override the forward time iteration process
 and attempt to identify the end configurations directly
 from the governing equations of the system.
 Although these end configurations have originated from
 dissipative reorganization through chaos,
 it is important to emphasize that these end configurations
 are determined by the ideal governing equations.
 Ideal MHD cannot, however, determine the dissipative path
 taken to reach one configuration from another.
 As a consequence, we are unable to recover the initial
 conditions by following the self-similar configurations
 backward in time. The initial conditions are lost from memory.

Recently, time self-similar analysis was applied to represent
 extragalactic jets as a polar-launched ejection event
 (Tsui and Serbeto 2007),
 which differs from the classical accretion-ejection space
 self-similar steady state MHD transport model
 (Blandford and Payne 1982).
 A self-similar description has been applied to different
 problems in physics, such as atmospheric ball lightnings
 (Tsui 2006, Tsui et al 2006), and interplanetary magnetic ropes
 (Osherovich et al 1993, 1995, Tsui and Tavares 2005).
 Here, we follow the methods of time self-similar MHD analysis
 to model plasma torus, by developing solutions that are launched
 onto the equatorial plane.

\newpage
\section{Axisymmetric Self-Similar MHD Scalings}

The basic MHD equations, in Eulerian fluid description,
 are given by
\\
$${\partial\rho\over\partial t}+\nabla\cdot(\rho\vec v)\,
 =\,0\,\,\,,\eqno(1)$$

$$\rho\{{\partial\vec v\over\partial t}
 +(\vec v\cdot\nabla)\vec v\}\,
 =\,\vec J\times\vec B-\nabla p
 -\rho{GM\over r^{3}}\vec r\,\,\,,\eqno(2)$$

$${\partial\vec B\over\partial t}\,
 =\,-\nabla\times\vec E\,
 =\,\nabla\times(\vec v\times\vec B)\,\,\,,\eqno(3)$$

$$\nabla\times\vec B\,=\,\mu\vec J\,\,\,,\eqno(4)$$

$$\nabla\cdot\vec B\,=\,0\,\,\,,\eqno(5)$$

$${\partial\over\partial t}({p\over\rho^{\gamma}})
 +(\vec v\cdot \nabla)({p\over\rho^{\gamma}})\,=\,0\,\,\,,\eqno(6)$$
\\
where $\rho$ is the mass density, $\vec v$ is the bulk velocity,
 $\vec J$ is the current density, $\vec B$ is the magnetic field,
 $p$ is the plasma pressure, $\mu$ is the free space permeability,
 $\gamma$ is the polytropic index, and $M$ is the central mass,
 which provides the gravitational field.

We consider a radially expanding plasma, and look for self-similar
 solutions in time, where time evolution is described by the
 dimensionless evolution function $y(t)$.
 For this purpose, it is most convenient to think of Lagrangian
 fluid description, and consider the position vector of a given
 laminar flow fluid element $\vec r(t)$.
 Under self-similarity, the radial profile is time invariant
 in terms of the radial label $\eta=r(t)/y(t)$,
 which has the dimension of $r$.
 Physically, $\eta$ is the Lagrangian
 radial position of a fixed fluid element.
 With a finite plasma, the domain of $\eta$ is bounded by mass
 conservation
\\
$$0\,<\,\eta_{int}\,<\,\eta\,<\,\eta_{ext}\,\,\,.$$
\\
As for the plasma velocity, we consider self-similar structures
 deriving from a spherically symmetric radial velocity that
 can be written as
\\
$$\vec v\,=\,{d\vec r(t)\over dt}\,
 =\{\eta{dy\over dt}, 0, 0\}\,
 =\,\{v, 0, 0\}\,\,\,.\eqno(7)$$
\\
Our self-similar parameter $\eta$, defined through
 the Lagrangian fluid label, explicitly represents
 the fluid velocity by the time evolution function $y(t)$.
 This evolution function will be solved self-consistently
 with respect to the spatial structures of the plasma.
 We emphasize that self-similarity, as a method, can be
 applied in different ways other than the one we use here.
 For example, Lou and his collaborators have treated an
 aggregating fluid under its self-gravitational field with a
 similarity variable $x=r(t)/at$, where $a$ is the sound speed,
 for isothermal fluid, $\gamma=1$,
 (Lou and Shen 2004, Bian and Lou 2005)
 and for a polytropic gas, $\gamma>1$,
 (Lou and Wang 2006, Lou and Gao 2006)
 to study relevant astrophysical phenomena.
 Extensions to a magnetofluid have been considered by Yu and Lou
 (2005) and by Lou and Wang (2007).
 Because of the linear dependence on time, this similarity
 variable $x$ refers to a reference frame moving at speed $a$,
 which is different from the radial plasma flow velocity $v$.
 Furthermore, different from our similarity variable $\eta$,
 $x$ here is not the Lagrangian label of a given fluid element.
 For this reason, the convective derivative remains explicit
 in the $x$ representation.
 As a result, the similarity variable of Lou amounts to finding
 the plasma structures in an adequate moving frame in the Eulerian
 $x$ fluid description.
 This resembles the analytic technique of going to a moving frame
 to look for stationary profile solutions for nonlinear phenomena
 such as nonlinear Alfven waves, solitons, etc.
 Because of this fundamentally different definition of
 self-similarity, the nature of the phenomena intended to
 describe is different.
 Our Lagrangian label self-similarity parameter is aimed to find
 spatial plasma configurations, and to determine the radial plasma
 flow velocity, consistent to the spatial configurations,
 through the evolution function.

Since we are considering an isotropic radial plasma flow,
 a natural solution would be a hydrodynamic one-dimensional
 expanding plasma, with radially dependent mass density $\rho$
 and plasma pressure $p$, and with $\vec B=0$ and $\vec J=0$.
 Nevertheless, this one-dimensional hydrodynamic solution
 is highly unlikely because magnetic field fluctuations
 can be generated from current density fluctuations,
 even in the absence of a pre-existing magnetic field.
 With the magnetic fields, which are basically a two- or
 three-dimensional structure, coupling to the plasma
 will generate likewise two- or three-dimensional $\rho$ and $p$.
 To examine the possible spatial structures of MHD plasmas,
 we consider a two-dimensional case with azimuthal symmetry in $\phi$.
 In this case, the magnetic field, through the vector potential
 $\vec A$, can be expressed as
\\
$$\vec B\,=\,{1\over r\sin\theta}
 \{+{1\over r}{\partial\over\partial\theta}(rA_{\phi}\sin\theta),
 -{\partial\over\partial r}(rA_{\phi}\sin\theta),
 +\sin\theta[{\partial\over\partial r}(rA_{\theta})
 -{\partial\over\partial\theta}(A_{r})]\}\,\,\,.$$
\\
This enables the magnetic field, the current density,
 and the electric field, to be expressed
 as two scalar functions $P$ and $Q$ respectively
\\
$$\vec B\,=\,{1\over r\sin\theta}
 \{+{1\over r}{\partial\over\partial\theta}(P),
   -{\partial\over\partial r}(P),+Q\}\,
 =\,\nabla P\times\nabla\phi+Q\nabla\phi\,\,\,,\eqno(8a)$$

$$\mu\vec J\,=\,{1\over r\sin\theta}
 \{+{1\over r}{\partial Q\over\partial\theta},
   -{\partial Q\over\partial r},
   -{\partial^2 P\over\partial r^2}
   -{1\over r^2}\sin\theta
   {\partial\over\partial\theta}
   ({1\over\sin\theta}{\partial P\over\partial\theta})\}
   \,\,\,,\eqno(8b)$$

$$\vec E\,=\,-(\vec v\times\vec B)\,
 =\,{1\over r\sin\theta}
 \{0, vQ, v{\partial P\over\partial r}\}\,\,\,.\eqno(8c)$$
\\
The independent variables are transformed from $(r,\theta,t)$
 space to $(\eta,\theta,y)$ space. We determine the explicit
 dependence of $y$ on each one of the physical variables
 with this radial velocity using functional analysis.
 First, making use of Eq.(7), Eq.(1) renders
\\
$${\partial\rho\over\partial t}
 +{1\over r^2}{\partial\over\partial r}(r^2v\rho)\,
 =\,({\partial\rho\over\partial t}
 +v{\partial\rho\over\partial r})
 +\rho({\partial v\over\partial r}+{2v\over r})\,
 =\,{\partial\rho\over\partial y}{dy\over dt}
 +{3\rho\over y}{dy\over dt}\,
 =\,({\partial\rho\over\partial y}+{3\rho\over y})
 {dy\over dt}\,=\,0\,\,\,.\eqno(9a)$$
\\
To reach the second equality, we note that the first bracket,
 in the first equality, corresponds to the total time derivative
 of an Eulerian fluid element, which amounts to the time derivative
 of a Lagrangian fluid element.
 As for the second bracket, it can be reduced by using
 $v=dr/dt=\eta dy/dt$ and $\partial v/\partial r=(1/y)(dy/dt)$.
 Solving this equation for $y$ scaling, by separating the time part,
 gives
\\
$$\rho(\vec r,t)\,
 =\,{1\over y^3}\bar\rho(\eta,\theta)\,\,\,.\eqno(9b)$$
\\
As for Eq.(6), with $\alpha_{0}F=(p/\rho^{\gamma})$ where $\alpha_{0}$
 is a constant that carries the physical dimension so that $F$ is
 a dimensionless function, it follows
\\
$${\partial F\over\partial t}+v{\partial F\over\partial r}\,
 =\,{\partial F\over\partial y}{dy\over dt}\,
 =\,0\,\,\,,\eqno(10a)$$

$$({p\over\rho^{\gamma}})\,=\,\alpha_{0}F(\vec r,t)\,
 =\,{1\over y^0}\alpha_{0}\bar F(\eta,\theta)\,\,\,.\eqno(10b)$$
\\
As for Eq.(3), using the representation of Eq.(8a), the magnetic
functions $P$ and $Q$ are
\\
$${\partial P\over\partial t}+v{\partial P\over\partial r}\,
 =\,{\partial P\over\partial y}{dy\over dt}\,
 =\,0\,\,\,,\eqno(11a)$$

$$P(\vec r,t)\,
 =\,{1\over y^0}\bar P(\eta,\theta)\,\,\,,\eqno(11b)$$

$${\partial Q\over\partial t}+{\partial\over\partial r}(vQ)\,
 =\,{\partial Q\over\partial y}{dy\over dt}
 +{Q\over y}{dy\over dt}\,
 =\,({\partial Q\over\partial y}+{Q\over y})
 {dy\over dt}\,
 =\,0\,\,\,,\eqno(12a)$$

$$Q(\vec r,t)\,
 =\,{1\over y^1}\bar Q(\eta,\theta)\,\,\,.\eqno(12b)$$
\\

\newpage
\section{Self-Similar Formulation}

We have reduced the general set of time-dependent ideal MHD
 equations, Eqs.(1-6), to a set of self-similar equations with
 appropriate time scalings, Eqs.(7-12), and have sought solutions
 of Eq.(2) and the time evolution function $y(t)$.
 The general ideal MHD set of equations has nonlinear terms
 of the convective type $(\vec v\cdot\nabla)$ in Eq.(1) and Eq.(2),
 and interaction type $(\vec J\times\vec B)$ in Eq.(2)
 and $(\vec v\times\vec B)$ in Eq.(3).
 By using the fluid label description, the $(\vec v\cdot\nabla)$
 convective terms are absorbed into the Lagrangian time derivative
 representation.
 The $(\vec v\times\vec B)$ interaction term is also absorbed,
 through an adequate representation of the magnetic field,
 by scalar functions $P$ and $Q$.
 The structure of the nonlinear terms,
 absorbed into the Lagrangian fluid label formulation,
 will appear in the $\eta$ profile of the system.
 The remaining task is to solve Eq.(2) for the self-similar $\eta$
 structure, under the presence of the $(\vec J\times\vec B)$
 interaction term.

We note that the equation of $\alpha_{0}F=(p/\rho^{\gamma})$ and
 the equation of $P$ are of the same form, and conclude that $F=F(P)$
 is a functional of $P$ only, or $\bar F=\bar F(\bar P)$.
 The fact that two equations have the same form does not
 automatically imply that the solution of one equation
 is a functional of the solution to the other.
 We can make this assertion only when considering self-similar solutions.
 We proceed to write the $(\eta,\theta)$ dependences
 in terms of $(\eta,\bar P(\eta,\theta))$ in $\bar\rho$ and
 $\bar p$, to obtain
\\
$$\bar\rho\,=\,\bar\rho(\eta,\bar P)\,\,\,,\eqno(13a)$$

$$p\,=\,{1\over y^{3\gamma}}\alpha_{0}\bar F(\eta,\theta)
 \bar\rho^{\gamma}(\eta,\theta)\, 
 =\,{1\over y^{3\gamma}}\alpha_{0}\bar F(\bar P)
 \bar\rho^{\gamma}(\eta,\bar P)\,
 =\,{1\over y^{3\gamma}}\bar p(\eta,\bar P)\,\,\,.\eqno(13b)$$
\\
Furthermore, since $(p/\rho^{\gamma})=\alpha_{0}F(P)$ is a function
 of $P$ only, the $\eta$ and $\bar P$ dependences should be in a
 separable form in both
\\
$$\bar\rho(\eta,\bar P)\,
 =\,\rho_{0}\bar\rho_{1}(\eta)\bar\rho_{2}(\bar P)\,\,\,,$$
 
$$\bar p(\eta,\bar P)\,
 =\,p_{0}\bar p_{1}(\eta)\bar p_{2}(\bar P)\,\,\,.$$
\\
The dimensions of mass density and pressure appear explicitly
 in $\rho_{0}$ and $p_{0}$ respectively. In this form,
 $\bar\rho_{1}(\eta)$, $\bar\rho_{2}(\bar P)$, $\bar p_{1}(\eta)$,
 and $\bar p_{2}(\bar P)$ are dimensionless functions.
 By taking $\alpha_{0}=p_{0}/\rho_{0}^{\gamma}$ with the physical
 dimension, we have
\\
$$\bar p_{1}(\eta)\,
 =\,\bar\rho_{1}^{\gamma}(\eta)\,\,\,,\eqno(14a)$$

$$\bar p_{2}(\bar P)\,
 =\,\bar F(\bar P)\bar\rho_{2}^{\gamma}(\bar P)\,\,\,,\eqno(14b)$$
\\
such that $\bar F$ should be a functional of $\bar P$ only.

Making use of Eq.(4) to eliminate the current density in Eq.(2),
 we derive the momentum equation, which has three components.
 The $\phi$ component, which contains only the magnetic force,
 gives
\\
$${\partial P\over\partial r}
 {\partial Q\over\partial\theta}
 -{\partial P\over\partial\theta}
 {\partial Q\over\partial r}\, 
 =\,{1\over y^2}
 \{{\partial\bar P\over\partial\eta}
 {\partial\bar Q\over\partial\theta}
 -{\partial\bar P\over\partial\theta}
 {\partial\bar Q\over\partial\eta}\}\,
 =\,0\,\,\,,\eqno(15)$$
 
$$\bar Q\,=\,\bar Q(\bar P)\,\,\,.\eqno(16)$$
\\
The function $P$ plus the functional form of $Q$, therefore,
 determines the magnetic fields. As for the $\theta$ component,
 with $p=\bar p(\eta,\bar P)/y^{3\gamma}$, it reads
\\
$${\partial^2 \bar P\over\partial\eta^2}
 +{1\over\eta^2}\sin\theta{\partial\over\partial\theta}
 ({1\over\sin\theta}{\partial\bar P\over\partial\theta})
 +\bar Q{\partial\bar Q\over\partial\bar P} 
 +\mu y^{4-3\gamma}\eta^2\sin^2\theta
 {\partial\bar p\over\partial\bar P}\,
 =\,0\,\,\,.\eqno(17)$$
\\
We remark that the first three terms of this equation represent the
 nonlinear force-free field equation with $\bar Q=\bar Q(\bar P)$.
 In the particular case of a linear functional $\bar Q=a\bar P$,
 this equation describes the linear force-free fields,
 which can be verified easily from Eqs.(8) with $\mu\vec J=a\vec B$.
 Such a force-free magnetic configuration is the marker of
 self-organized plasmas, and is the result of the variational
 principle that minimizes the energy density of the magnetic field,
 under the constraint of global magnetic helicity conservation.
 This justifies the use of self-similarity method to describe
 self-organized configurations.
 The last term of Eq.(17) is the plasma pressure term.
 For an isotropic pressure independent of $\bar P$, this last
 term would be null and such plasma pressure $\bar p(\eta)$
 has no effect on the magnetic field which remains force-free.
 For an axisymmetric pressure $\bar p(\eta,\bar P)$,
 the above equation would be independent of the time evolution
 function $y$ should $\gamma=4/3$, and it would describe
 a plasma pressure balanced axisymmetric magnetic field.
 Such a pressure-balanced self-organized configuration could be
 derived using the variational principle,
 by minimizing the energy density that contains
 not only magnetic energy but also plasma thermal energy
 under the same constraint.
 The solution of this equation of $P$, with different representations
 of $\bar Q=\bar Q(\bar P)$ and $\bar p=\bar p(\bar P)$, is the core
 problem of self-similar description.

As for the $r$ component of the momentum equation, it reads
\\
$$\rho[{\partial v\over\partial t}
 +v{\partial v\over\partial r}]
 +{d\over dr}p(r,P(r,\theta))
 +\rho{GM\over r^2}\,$$
 
$$=\,-{1\over\mu}({1\over r\sin\theta})^2
 {\partial P\over\partial r}
 \{{\partial^2 P\over\partial r^2} 
 +{1\over r^2}\sin\theta{\partial\over\partial\theta}
 ({1\over\sin\theta}{\partial P\over\partial\theta})
 +Q{\partial Q\over\partial P}\}\, 
 =\,{\partial P\over\partial r}{\partial \over\partial P}p(r,P)
 \,\,\,.\eqno(18a)$$
\\
The term $dp/dr$ on the left side refers to the total radial
 derivative of the plasma pressure,
 which includes both an explicit and implicit dependence in $P$.
 The right side of the radial equation provides an expression
 for the magnetic force.
 Making use of the meridian component, Eq.(17), the right side
 is equal to the implicit part of the radial pressure gradient,
 which cancels the same term on the left side.
 This leaves only the explicit radial pressure gradient
\\
$$\rho[{\partial v\over\partial t}
 +v{\partial v\over\partial r}]
 +{\partial \over\partial r}p(r,P)
 +\rho{GM\over r^2}\, 
 =\,\rho{d^2r\over dt^2}
 +{\partial \over\partial r}p(r,P)
 +\rho{GM\over r^2}\,
 =\,0\,\,\,.\eqno(18b)$$
\\
In terms of self-similar parameters, the radial equation reads
\\
$$y^2{d^2y\over dt^2}
 +{1\over\eta}
 \{y^{4-3\gamma}{1\over\bar\rho}
 {\partial\bar p\over\partial\eta}
 +{GM\over\eta^2}\}\,
 =\,0\,\,\,.\eqno(19)$$
\\
With $\alpha$ as the separation constant, $\alpha_{2}$ as a
 dimensionless constant, and $\gamma=4/3$, we derive
\\
$$4\alpha_{2}{p_{0}\over\rho_{0}}\bar\rho_{1}^{1/3}(\eta)\,
 =({1\over 2}\alpha\eta^{2}+{GM\over\eta})\, 
 ={\alpha\over 2\eta}(\eta^{3}+{2GM\over\alpha})\,
 ={\alpha\over 2\eta}(\eta^{3}+2\eta_{*}^{3})
 \,\,\,,\eqno(20a)$$

$$\bar p_{2}(\bar P)\,
 =\,\alpha_{2}\bar\rho_{2}(\bar P)\,\,\,,\eqno(20b)$$

$$y^2{d^2y\over dt^2}\,=\,-\alpha\,\,\,.\eqno(20c)$$
\\
The first of these three equations provides the mass density
 profile of the radial label. The second one specifies the
 $\bar P$ dependence between mass density and plasma pressure.
 The third one defines the time history of the evolution function.

\newpage
\section{Axisymmetric Equatorial Structures}

After describing our self-similar formulation, we solve Eqs.(14),
 (17), and (20), for axisymmetric plasma configurations.
 We begin with comparing Eqs.(14b) and (20b), and conclude that
\\
$$\bar F(\bar P)\,
 =\,\alpha_{2}\bar\rho_{2}^{1-\gamma}(\bar P)\,\,\,.\eqno(21)$$
\\
The solution to Eq.(17), for the magnetic field function $\bar P$,
 depends on the functionals $\bar Q$ and $\bar p_{2}$.
 This is where different types of self-similar configurations
 can be produced.
 Although Low pioneered the use of self-similar MHD in astrophysics,
 he did not explore the full potential of the method.
 We remark that self-similar solutions are highly dependent on
 the functional representations of $\bar Q$ and $\bar p_{2}$.
 Under a given geometry, there are different self-similar solutions,
 with different characteristics to represent different self-organized plasmas,
 depending on how the magnetic fields and plasma parameters are modeled.
 To study configurations on and about the equatorial plane, we consider
\\
$$\bar Q\,=\,a\bar P\,\,\,,\eqno(22a)$$

$$\bar p_{2}(\bar P)\,=\,(\bar P+\bar C)\,>\,0\,\,\,.\eqno(22b)$$
\\
For example, if we assume a quadratic function to have
\\
$$\bar p_{2}(\bar P)\,=\,(\bar P^{2}+\bar C)\,>\,0\,\,\,,$$
\\
we derive a polar ejection configuration that tends
 to collimate magnetic fields and plasma density, onto the polar axis,
 as plasma pressure builds up (Tsui and Serbeto 2007).
 Using these functional representations in Eq.(17), we derive
\\
$$\eta^{2}{\partial^2\bar P\over\partial\eta^2}
 +\sin\theta{\partial\over\partial\theta}
 ({1\over\sin\theta}{\partial\bar P\over\partial\theta})
 +(a\eta)^{2}\bar P\,$$
 
$$ =\,-\mu p_{0}\eta^{4}\bar p_{1}(\eta)\sin^2\theta\,
 =\,-\mu p_{0}\eta^{4}\bar\rho_{1}^{4/3}(\eta)\sin^2\theta
 \,\,\,.\eqno(23)$$
\\
To solve this equation, we separate the variables by writing
 $\bar P(\eta,\theta)=R(\eta)\Theta(\theta)$.
 Assuming $x=\cos\theta$ and $n(n+1)$ to be the separation constant,
 we have
\\
$$(1-x^2){d^2\Theta(x)\over dx^2}
 +n(n+1)\Theta(x)\,=\,0\,\,\,,\eqno(24a)$$

$$\eta^{2}{d^2R\over d\eta^2}
 +[(a\eta)^{2}-n(n+1)]R\, 
 =\,-\mu p_{0}\eta^{4}\bar\rho_{1}^{4/3}(\eta)
 {\sin^2\theta\over\Theta(x)}\,\,\,.\eqno(24b)$$
\\ 
The first equation provides the well-known solution,
 in terms of the Legendre polynomial $P_{n}(x)$,
\\
$$\Theta(x)\,=\,C_{0}(1-x^2){dP_{n}(x)\over dx}\,
 =\,-n(n+1)C_{0}\int_{1}^{x}P_{n}(x)dx\,\,\,.\eqno(25)$$
\\
We note that the second equation is not in a separable form.
 Nevertheless, should we choose $n=1$ with $P_{1}(x)=x$,
 such that $\Theta(x)=(1-x^2)$, where we assume that $C_{0}=1$,
 Eq.(24b) is separable.
 The general solution is given by a homogeneous solution
 $R_{0}(\eta)$ and a particular solution $R_{1}(\eta)$
\\
$$R(\eta)\,=\,R_{0}(\eta)+R_{1}(\eta)\,
 =\,A_{0}(a\eta)j_{n}(a\eta)+R_{1}(\eta)\,\,\,,\eqno(26)$$
\\
where $j_{n}(a\eta)$ with $n=1$ is the spherical Bessel function
 regular at $\eta=0$.
 
The coefficient $A_{0}$ of the homogeneous solution carries the
 amplitude and dimension to reflect the magnetic field through Eq.(8a).
 Since plasma pressure and mass density are expressed in terms of
 $P$ via $\bar p_{2}(\bar P)$ and $\bar\rho_{2}(\bar P)$,
 we could normalize the entire self-similar solution,
 with respect to the magnetic field, by taking $A_{0}=1$.
 With $\bar\rho_{1}(\eta)$ given by Eq.(20a), denoting $z=a\eta$,
 and $\kappa^{3}=2(a\eta_{*})^{3}=2z_{*}^{3}$,
 the particular solution is described by
\\
$$z^{2}{d^2R_{1}\over dz^2}
 +[z^{2}-n(n+1)]R_{1}\, 
 =\,-\mu p_{0}({\alpha\over a^3})^{4}
 ({1\over 8\alpha_{2}}{\rho_{0}\over p_{0}})^{4}
 (z^{3}+\kappa^{3})^{4}\,
 =\,-A(z^{3}+\kappa^{3})^{4}\,\,\,.$$
\\
We note that the coefficient $A$, which has two factors, is dimensionless.
 The first factor $({\mu p_{0}/a^4})\simeq \beta/z_{*}^{2}$ is related
 to the plasma $\beta$,
 which is the ratio of plasma pressure to magnetic pressure.
 The second dimensionless factor with $GM$ is basically the ratio of
 the gravitational potential energy density to the plasma pressure.
 The right side of this equation can be expanded into binomial terms,
 and can be solved using power series, by taking one term at a time.
 The particular solution is, therefore, a superposition
 of five sub-solutions $R_{1(12)}$, $R_{1(9)}$, $R_{1(6)}$,
 $R_{1(3)}$, and $R_{1(0)}$,
 where the bracketed number in the subscript denotes the power
 of $z$ in the binomial term, on the right side.
 We denote the coefficients $A^{(12)}=A$,
 $A^{(9)}=4\kappa^{3}A$, $A^{(6)}=6\kappa^{6}A$,
 $A^{(3)}=4\kappa^{9}A$, $A^{(0)}=\kappa^{12}A$.
 We solve each binomial term of the particular solution
 using a power series of the form
\\
$$R_{1(k)}(\eta)\,=\,\sum a_{m}z^{+m}\,\,\,.$$
\\
Using standard power-series techniques,
 we derive solutions of the form
\\
$$R_{1(12)}(z)\,
 =\,a_{2}z^{2}+a_{4}z^{4}+a_{6}z^{6}+a_{8}z^{8}
 +a_{10}z^{10}\,\,\,,\eqno(27a)$$
\\
where $a_{10}=-A^{(12)}$, $a_{8}=-(10\times 9-2)a_{10}$,
 $a_{6}=-(8\times 7-2)a_{8}$, $a_{4}=-(6\times 5-2)a_{6}$,
 $a_{2}=-(4\times 3-2)a_{4}$, and $a_{0}=-(2\times 1-2)a_{2}=0$,
\\
$$R_{1(9)}(z)\,
 =\,a_{9}z^{9}+a_{11}z^{11}+a_{13}z^{13}+a_{15}z^{15}
 + ..... \,\,\,,\eqno(27b)$$
\\
where $a_{9}=-A^{(9)}/(9\times 8-2)$, $a_{11}=-a_{9}/(11\times 10-2)$,
 $a_{13}=-a_{11}/(13\times 12-2)$, $a_{15}=-a_{13}/(15\times 14-2)$,
 $a_{m}=-a_{m-2}/(m\times (m-1)-2)$,
\\
$$R_{1(6)}(z)\,
 =\,a_{2}z^{2}+a_{4}z^{4}\,\,\,,\eqno(27c)$$
\\
where $a_{4}=-A^{(6)}$, $a_{2}=-(4\times 3-2)a_{4}$,
 and $a_{0}=-(2\times 1-2)a_{2}=0$,
\\
$$R_{1(3)}(z)\,
 =\,a_{3}z^{3}+a_{5}z^{5}+a_{7}z^{7}+a_{9}z^{9}
 + ..... \,\,\,,\eqno(27d)$$
\\
where $a_{3}=-A^{(3)}/(3\times 3-2)$, $a_{5}=-a_{3}/(5\times 4-2)$,
 $a_{7}=-a_{5}/(7\times 6-2)$, $a_{9}=-a_{7}/(9\times 8-2)$,
 $a_{m}=-a_{m-2}/(m\times (m-1)-2)$.
 As for $R_{1(0)}(z)$, this solution has to be solved with series
 of negative powers, which describes the divergent nature of the
 gravitational field $GM/\eta$ at $\eta=0$,
 although this limit is irrelevant here.
 We derive
\\
$$R_{1(0)}(\eta)\,=\,\sum a_{m}z^{-m}\,\,\,,\eqno(27e)$$
\\
where $a_{0}=0$, $a_{2}=-A^{(0)}$,
 $a_{4}=-(2\times 3-2)a_{2}/(4\times 5)$,
 $a_{6}=-(4\times 5-2)a_{4}/(6\times 7)$, 
 $a_{8}=-(6\times 7-2)a_{6}/(8\times 9)$,
 $a_{m}=-((m-2)\times (m-1)-2)a_{m-2}/(m\times (m+1))$.
 We note that the homogeneous solution corresponds to force-free
 magnetic field configurations. Due to the plasma pressure in Eq.(17),
 there are particular solutions that maintain the pressure balance.

\newpage
\section{Magnetic Fields and Evolution Function}

After solving for the spatial structure of $\bar P$,
 the magnetic fields are given by
\\
$$B_{r}\,=\,{1\over y^2}\bar B_{r}(\eta,\theta)\,
 =\,+{1\over y^2}{1\over\eta\sin\theta}
 {1\over\eta}{\partial\bar P\over\partial\theta}\, 
 =\,-{1\over y^2}{1\over\eta^2}R(\eta)
 {d\Theta(x)\over dx}\,\,\,,\eqno(28a)$$ 

$$B_{\theta}\,=\,{1\over y^2}\bar B_{\theta}(\eta,\theta)\,
 =\,-{1\over y^2}{1\over\eta\sin\theta}
 {\partial\bar P\over\partial\eta}\, 
 =\,-{1\over y^2}{1\over\eta}{dR(\eta)\over d\eta}
 {1\over (1-x^2)^{1/2}}\Theta(x)\,\,\,,\eqno(28b)$$ 

$$B_{\phi}\,=\,{1\over y^2}\bar B_{\phi}(\eta,\theta)\,
 =\,+{1\over y^2}{1\over\eta\sin\theta}\bar Q\, 
 =\,+{1\over y^2}{a\over\eta}R(\eta)
 {1\over (1-x^2)^{1/2}}\Theta(x)\,\,\,.\eqno(28c)$$ 
\\
The self-similar homogeneous solution $R_{0}(\eta)$,
 provided by Eq.(26),
 enables oscillations in $\eta$ to be modeled,
 and vanishes at the zeros of the spherical Bessel function.
 To understand the magnetic structures,
 we first set aside the particular solution $R_{1}(\eta)$.
 At the first zero of $R_{0}(\eta)$, we have
 $B_{\phi}(\eta)=0$, $B_{r}(\eta)=0$, and the only nonvanishing
 field is $B_{\theta}(\eta)=0$ according to Eqs.(28).
 The meridian self-similar solution $\Theta(x)$,
 given by Eq.(25), oscillates in $x$.
 Together they describe the magnetic fields of Eqs.(28).
 Within this region of $(\eta,x)$, the topological center,
 defined by $dR_{0}(\eta)/d\eta=0$ and $d\Theta(x)/dx=0$,
 has $B_{r}=0$, $B_{\theta}=0$, and the only nonvanishing
 field is $B_{\phi}(\eta)=0$.
 This is the magnetic axis.
 The field lines about the magnetic axis are, therefore, given by
\\
$${B_{r}\over dr}\,=\,{B_{\theta}\over rd\theta}\,
 =\,{B_{\phi}\over r\sin\theta d\phi}\,\,\,.$$
\\
By axisymmetry, the third group is decoupled from the first two groups.
 For the field lines on an $(r-\theta)$ plane,
 we consider the first equality between $B_{r}$ and $B_{\theta}$,
 which gives
\\
$$\bar P(\eta,x)\,=\,R_{0}(\eta)(1-x^2)\,=\,C\,\,\,.\eqno(29a)$$
\\
The nested field lines are given by the contours of
 $\bar P(\eta,x)$ on the $(r-\theta)$ plane.
 At the topological center, the magnetic axis,
 we have $\Theta(x)$ at its maximum value,
 and $R_{0}(\eta)$ at its maximum,
 such that $\bar P(\eta,x)$ takes its maximum value.
 Since
\\
$$2\pi r\sin\theta B_{\phi}\,=\,2\pi a\bar P\,\,\,,\eqno(29b)$$
\\
by Eq.(28c), the line integral of $B_{\phi}$ on the magnetic axis
 is a maximum.
 We, therefore, have a sequence of plasma tori,
 due to the periodic nature of the spherical Bessel function.
 The presence of the particular solution $R_{1}(\eta)$ modifies
 this structure.
 Due to the divergent nature of $R_{1}(\eta)$,
 as we will see in the next section,
 only the first torus prevails.

As for the temporal part, the evolution function in Eq.(20c)
 is described by
\\
$${1\over 2}({dy\over dt})^2\,
 =\,(H+{\alpha\over y})\,>\,0\,\,\,,\eqno(30a)$$

$$({dy\over dt})
 =\,[2(H+{\alpha\over y})]^{1/2}\,\,\,.\eqno(30b)$$
\\
Here, $H$ is an integration constant that is independent of time.
 To understand the meaning of $\alpha$, we note that,
 using Eq.(20c), plasma acceleration in Lagrangian coordinates is
\\
$${dv\over dt}\,=\,\eta{d^2y\over dt^2}\,
 =\,-{\alpha\eta\over y^2}\,\,\,.\eqno(31a)$$
\\
A positive $\alpha$ corresponds to an outward, decelerating flow.
 The deceleration becomes smaller as $y$, or as $r$, becomes larger,
 and $\alpha$ equals the intensity of the deceleration,
 for a given radial label $\eta$.
 To derive an expression for $H$, we multiply Eq.(30a) by $\rho\eta^2$,
 and use Eq.(31a) to obtain
\\
$$\rho\eta^{2}H\,
 =\,{1\over 2}\rho v^{2}+(\rho{dv\over dt})r\, 
 =\,{1\over 2}\rho v^{2}
 -({\partial \over\partial r}p(r,P)
 +\rho{GM\over r^2})r\,\,\,.\eqno(31b)$$
\\
The first term on the right side is the kinetic energy density of
 the fluid element, and the second term is its work on expansion,
 due to the explicit part of the pressure gradient.
 We note that the magnetic energy does not appear in this expression
 because it is cancelled by the work due to the implicit pressure
 gradient, as discussed in Eq.(18a).
 It is clear that $H$ measures the total energy of the fluid element.

\newpage
\section{Force-Free Equatorial Plasma Torus}

We consider the explicit radial dependent part of the mass
 density, in Eq.(20a). Since $\bar\rho_{1}(\eta)$ is positive,
 $\alpha$ should also be positive. Differentiating with respect to
 $\eta$ indicates that $\bar\rho_{1}(\eta)$ has a minimum at
\\
$$\eta_{*}^{3}\,=\,{GM\over \alpha}\,\,\,.\eqno(32)$$
\\
Because we are considering a finite plasma, the range of the radial
 label $\eta$ is finite.
 We assume that $\eta_{*}$ is enclosed in the interval
 $0<\eta_{int}<\eta_{*}<\eta_{ext}$.
 For $\eta>\eta_{*}$, $\bar\rho_{1}(\eta)$ increases
 because of the $\alpha\eta^{2}/2$ term,
 where $\alpha$ is defined by Eq.(20c)
 and measures the inertia of the plasma.
 A positive $\alpha$ corresponds to a decelerating plasma.
 Beyond the external bound is interstellar space,
 where plasma mass density decreases abruptly.
 In the presence of the gravitational term, $GM/\eta$, in Eq.(20a),
 the mass density $\bar\rho_{1}(\eta)$ rises again for $\eta<\eta_{*}$.
 This gravitational term is singular at $\eta=0$.
 However this singularity is not included,
 because the internal bound of $\eta$ is at the surface of the star.
 
To consider the function $\bar P(\eta,x)=R(\eta)\Phi(x)$,
 the homogeneous solution $R_{0}(\eta)=A_{0}(a\eta)j_{1}(a\eta)$,
 with $A_{0}=1$, is shown in Fig.1,
 with the first node at $z=(a\eta)=4.5$.
 In the absence of plasma pressure, $R_{0}(\eta)$ would be the only
 term for $R(\eta)$. With $z_{ext}=a\eta_{ext}=4.5$, this would
 correspond to the equatorial plasma torus force-free magnetic fields.
 As for the particular solution $R_{1}(\eta)$, we note that
 $R_{1(12)}(z)$, $R_{1(9)}(z)$, $R_{1(6)}(z)$, and $R_{1(3)}(z)$
 diverge as $z$ goes to infinity,
 whereas $R_{1(0)}(z)$ diverges as $z$ goes to zero.
 These two extremes are naturally excluded,
 because $z=0$ is the center of the central star,
 and $z=\infty$ requires an infinite plasma, which is unphysical.
 An adequate domain could be provided by the homogeneous solution
 plotted in Fig.1,
 which shows an axisymmetric equatorial plasma torus with $0<z<4.5$,
 if the particular solution does not significantly alter this structure.
 The particular binomial solutions depend on the
 parameter $\kappa^{3}=2GMa^{3}/\alpha=2(a\eta_{*})^{3}$,
 where $a$ is a measure of the azimuthal magnetic field given by Eq.(22a).
 We assume the minimum $z_{*}=a\eta_{*}=2$, such that $\kappa^{3}=16$.
 We note that the two factors (brackets) in
\\
$$A\,=\,-({\mu p_{0}\over a^4})
 ({1\over 8\alpha_{2}}{1\over z_{*}^{2}}
 {\rho_{0}GM\over\eta_{*}}{1\over p_{0}})^{4}\,\,\,$$
\\
in the equation of $R_{1}(\eta)$ are dimensionless quantities.
 The first factor
 $(\mu p_{0}/a^{4})\simeq \beta/z_{*}^{2}\simeq 10^{-1}$
 can be expressed in terms of plasma $\beta$,
 which is the ratio of the plasma pressure to the magnetic pressure.
 This factor is estimated to be $10^{-1}$.
 The second factor is basically the ratio of the gravitational potential
 energy density $\rho_{0}GM/\eta_{*}$ to the plasma pressure $p_{0}$,
 plus other dimensionless multipliers to the fourth power.
 We believe that the gravitational potential energy density
 and the plasma pressure are approximately similar.
 We could take $\alpha_{2}=1$, and the remaining terms
 would provide the factor $10^{-6}$.
 Multiplying the two factors together gives $A=10^{-7}$.

Of the five parts of the particular solution $R_{1}$,
 $R_{1(12)}$ is unaffected by the choice of $\kappa^{3}$,
 and is plotted in Fig.2.
 Other binomial parts $R_{1(9)}$, $R_{1(6)}$, $R_{1(3)}$,
 plotted in Figs.3-5 respectively,
 carry mixed contributions of plasma pressure and gravitational field,
 and are negligible within the domain.
 The last part $R_{1(0)}$ in Fig.6 is purely gravitational.
 The neighborhood of the center of the star, $z=0$, is excluded.
 These binomial parts depend on the choice of $\kappa^{3}$,
 but would have a negligible effect on $R_{0}$,
 within the domain of interest.
 The homogeneous and the particular solutions
 provide the complete solution for $R(\eta)$.
 Together with $\Theta(x)$, they determine the function $\bar P(\eta,x)$,
 which governs the entire axisymmetric self-similar profile
 in magnetic fields, plasma pressure, and mass density.

To consider the magnetic field profiles, we recall that the field
 lines on an $(r-\theta)$ plane are given by Eq.(29a),
 with $R(z)=R_{0}(z)+R_{1}(z)$.
 The presence of a plasma gives a diamagnetic effect,
 as shown by $R_{1(12)}(z)$ in Fig.2,
 carrying an opposite sign with respect to $R_{0}(z)$ of Fig.1.
 Examining Fig.2, we note that $R_{1(12)}$ vanishes at $z=4.5$
 and $z=5.2$.
 The first node of $R_{1(12)}$ happens to be at the
 same location as the first node of $R_{0}$, also at $z=4.5$.
 Furthermore, Figs.(3-6) show that $R_{1(9)}$, $R_{1(6)}$,
 $R_{1(3)}$, $R_{1(0)}$ are all negligibly small.
 Consequently, the general solution $R(z)=R_{0}(z)+R_{1}(z)=0$
 at $z=4.5$.
 By Eqs.(28), we conclude that the radial and azimuthal
 magnetic fields, $B_{r}(z)$ and $B_{\phi}(z)$,
 vanish at this location, while the meridian magnetic field,
 $B_{\theta}(z)$, is maximum.
 Consequently, the domain $0<z<4.5$ constitutes a region for a
 magnetic torus with closed field lines.
 Since plasma is tied to the magnetic field lines,
 it circulates along and is confined within this torus.
 We note that the sign of $R_{1}$ is negative,
 which indicates the diamagnetic nature of plasma.
 We note that $R_{1(12)}$ has a maximum at the same location as $R_{0}$.
 As a matter of fact, the functional form of $R_{1(12)}$ in the
 interval $(0,4.5)$ is much the same as that of the force-free $R_{0}$.
 Consequently, the general solution $R$ is practically force-free,
 even when plasma pressure is taken into account, in Eq.(23).
 
For simplicity, we neglect the diamagnetic effect,
 and take $R(z)=R_{0}(z)$ to highlight the poloidal field lines in Fig.7,
 which is described by first equality of the field line equation,
 with solutions in Eq.(29a).
 The addition of $R_{1(12)}(z)$ would change the contours slightly,
 but not the topology.
 Apart from these poloidal fields,
 there are associated toroidal fields given by Eq.(28c).
 The maxima of $R$ and $\Theta$, ${\partial\bar P/\partial\eta}=0$
 and ${\partial\bar P/\partial\theta}=0$, give the maximum of this
 toroidal field where $B_{r}=0$ and $B_{\theta}=0$ respectively.
 This corresponds to the magnetic axis of the plasma torus.
 Together, the field lines thread out nested magnetic surfaces
 enclosing the magnetic axis, one within another.
 The plasma pressure and mass density of this plasma torus have
 $R(z)>0$. With $\Theta(x)>0$, this warrants $\bar P(z,x)>0$
 to assure positive pressure and mass density.
 With $\bar C=0$ in Eq.(22b), the plasma pressure and mass density
 vanish at $z=4.5$ since $R(z)=0$.
 It is important to note that, for $z>4.5$, the particular solution
 $R_{1}(z)$ dominates the homogeneous solution $R_{0}(z)$ such that
 the general solution $R(z)$ never vanishes.
 The radial component of the magnetic field is nonzero.
 For this reason, the field lines are open to outer space,
 and the plasma is no longer confined.
 This outer open plasma structure is not shown in Fig.7.

For the evolution function, it is most convenient to write Eq.(30b)
 in the normalized form
\\
$$({dy\over d\tau})
 =\,[{1\over y}\pm {|H|\over\alpha}]^{1/2}\,\,\,,\eqno(33)$$
\\
with $\tau=(2\alpha)^{1/2}t$ as the normalized time.
 We assume that $\alpha$ is positive, and take $y(0)=1$
 to be the initial value, such that the Lagrangian label $\eta$
 corresponds to the initial position of the Eulerian fluid element.
 Equation (33) indicates an asymptotically convergent solution,
 with $H=-|H|<0$ negative, which is shown in the lower curve
 of Fig.8 with $|H|/\alpha=0.5$.
 This track represents an asymptotically stable radius
 for the equatorial plasma torus.
 On the other hand, if the energy density $H=+|H|>0$ is positive,
 there is an ever expanding evolution track with a terminal
 velocity $dy/d\tau=(H/\alpha)^{1/2}$.
 This is shown in the upper curve of Fig.8,
 which is inappropriate for AGB slow winds.
 
To better understand the amplitude and the sign of $H$,
 we multiply Eq.(30a) by $2\rho\eta d\eta$,
 and integrate to derive
\\
$$2H\int \rho\eta\,d\eta\,
 =\,({dy\over dt})^2\int\rho\eta\,d\eta 
 -2\int({\partial p\over\partial r}+\rho{GM\over r^2})\,dr
 \,\,\,.\eqno(34)$$
\\
The first integral on the right side is positive definite,
 and corresponds to the kinetic energy of the expanding structure.
 Inside the second integral, there is a plasma pressure term,
 and a gravitational term.
 The gravitational term is always positive.
 Within the plasma pressure term, we have
 $\partial \bar p_{1}(\eta)/\partial\eta<0$ for $\eta<\eta_{*}$,
 since $\bar p_{1}(\eta)$ has a similar profile as
 $\bar\rho_{1}(\eta)$, according to Eq.(14a).
 This part has a forward plasma pressure force.
 For $\eta>\eta_{*}$, we have
 $\partial \bar p_{1}(\eta)/\partial\eta>0$, and the corresponding
 plasma pressure force is backward.
 It is clear that the explicit part of the pressure profile
 $\bar p_{1}(\eta)$,
 which is described by the explicit part of the mass density
 profile $\bar\rho_{1}(\eta)$ in Eq.(14a),
 contributes to the sign of $H$.
 Rewriting Eq.(20a) as
\\
$$\bar\rho_{1}^{1/3}(\eta)\,
 =\,{1\over 4\alpha_{2}}{\rho_{0}\over p_{0}}\alpha\eta^{2}_{*}
 [{1\over 2}({\eta\over\eta_{*}})^{2}+({\eta_{*}\over\eta})]
 \,\,\,,\eqno(35)$$
\\
and assuming that the coefficient on the right side has a value of unity,
 these two profiles are shown in Fig.9,
 where we have taken $\eta=3\eta_{*}$ or $z=3z_{*}=6$
 as the upper bound in Fig.9, as an example.
 The pressure profile in the positive gradient region $\eta>\eta_{*}$
 helps to cancel the effects of the negative gradient,
 over the region $\eta<\eta_{*}$, in the above equation.
 Assuming that the slow wind velocity is below the escape velocity,
 the first integral on the right side is less than the gravitational
 term of the second integral.
 This implies that $H$ is negative,
 due to the overall negative sign on the second integral.
 The plasma torus is, therefore, asymptotically stationary.

When deriving the full radial profile of the plasma pressure,
 we caution that Fig.9 is only the explicit part,
 $\bar p_{1}(\eta)$, of the radial profile relevant to Eq.(34).
 Another radial contribution comes from the implicit part,
 $\bar p_{2}(\bar P)$.
 The full radial pressure profile is a combination of the
 explicit and the implicit part,
 $\bar p(\eta,\bar P)=p_{0}\bar p_{1}(\eta)\bar p_{2}(\bar P)$,
 where $p_{2}(\bar P)=\bar P(\eta,x)=R(\eta)\Theta(x)$.
 Since $\bar p_{1}(\eta)=\bar\rho_{1}^{\gamma}(\eta)$ by Eq.(14a)
 and $\bar p_{2}(\bar P)=\alpha_{2}\bar\rho_{2}(\bar P)$ by Eq.(20b),
 we could plot the full mass density profile,
 which reflects the plasma pressure profile with $\bar C=0$, as
\\
$$\bar\rho(\eta,\bar P)\,
 =\,\rho_{0}\bar\rho_{1}(\eta)\bar\rho_{2}(\bar P)\, 
 =\,\rho_{0}[{1\over 4\alpha_{2}}{\rho_{0}\over p_{0}}
 {\alpha z_{*}^2\over a^2}]^{3}
 [{1\over 2}({z\over z_{*}})^{2}+({z_{*}\over z})]^{3}
 {1\over\alpha_{2}}R(\eta)\Theta(x)\,\,\,.\eqno(36)$$
\\
With $z_{*}=2$ and considering $R(\eta)=R_{0}(\eta)$,
 the full radial dependence $\bar\rho_{1}(\eta)R_{0}(\eta)$,
 with the value of the coefficient in Eq.(36) taken to be unity,
 is shown in Fig.10 for the plasma torus with a very pronounced
 maximum at $z=3.7$, and zero at $z=4.5$.
 The corresponding mass density spatial contours
 of $\bar\rho(\eta,\bar P)$ are shown in Fig.11.

\newpage
\section{Bipolar Planetary Nebulae}

In addition to the equatorial plasma torus, our self-similar MHD
 formulation is able to model planetary nebulae of the bipolar type.
 For this purpose, we consider
\\
$$\bar Q\,=\,0\,\,\,,\eqno(37)$$
\\
to warrant $B_{\phi}=0$. Following the same analysis, the radial
 function becomes
\\
$$R(\eta)\,=\,R_{0}(\eta)+R_{1}(\eta)\,
 =\,A_{0}z^{-1}
 -a_{12}z^{12}-a_{9}z^{9}-a_{6}z^{6}-a_{3}z^{3}+a_{0}
 \,\,\,.\eqno(38)$$
\\
The coefficients are $(12\times 11-2)a_{12}=A^{(12)}=A$,
 $(9\times 8-2)a_{9}=A^{(9)}=4\kappa^{3}A$,
 $(6\times 5-2)a_{6}=A^{(6)}=6\kappa^{6}A$,
 $(3\times 2-2)a_{3}=A^{(3)}=4\kappa^{9}A$,
 and $2a_{0}=A^{(0)}=\kappa^{12}A$.
 Although we have continued to use $z=a\eta$ as the normalized
 radial label, the normalization factor $a$ here has no
 connection to $\bar Q$ because we now assume that $\bar Q=0$.
 We could imagine that $a=k$, such that $z=k\eta$.
 We notice that the particular solution consists entirely
 of negative terms, apart from the final term,
 whereas the homogeneous solution is positive.
 With $A_{0}=1$, $A=10^{-7}$, $z_{*}=2$,
 $\kappa^3=2z_{*}^3=16$, the general solution $R(\eta)$
 shown in Fig.12 becomes negative at $z=z_{0}=4.55$.
 This value of $Z_{0}$ happens to be close to the roots of
 $R_{0}(z)$ and $R_{1}(z)$,
 of the equatorial plasma torus case,
 where $\bar Q=a\bar P\neq 0$.
 This fact is purely a coincidence.
 The full radial profile of the plasma density,
 described by Eq.(36),
 explicit and implicit parts, is shown in Fig.13.
 Since plasma density is positive definite, $z\leq z_{0}$
 defines the physical domain of our self-similar solution.
 The bipolar field contours are shown in Fig.14.
 The radial component vanishes at $z_{0}$, $B_{r}(z_{0})=0$,
 which is the boundary of the expanding bipolar structure.

To conclude, we have represented the AGB wind by an exploding
 ideal MHD plasma. The explosion is modelled by an isotropic radial
 plasma velocity in spherical coordinates.
 Through dissipation,
 this exploding plasma is believed to undergo self-organization.
 The possible self-organized states may or may not be spherically
 symmetric, although the velocity will be spherical symmetric.
 Self-organization implies that the plasma global conservation
 properties are sufficiently dominant to drive the system
 through dissipation, independently of the initial conditions.
 To find the possible self-organized states, we solve the ideal MHD
 equations, with $\gamma=4/3$, for self-similar solutions in time,
 where the temporal and spatial dependences of each physical
 variable are in separable form.
 We identify these self-similar solutions of ideal MHD to be
 the self-organized configurations reached by dissipation.
 We would like to emphasize the word \textbf{ideal},
 which means that this process occurs without dissipation.
 Although self-organized states are physically generated
 by dissipation, they are part of the ideal MHD solutions.
 Dissipation provides the means by which an ideal state,
 of a particular energy density, is topologically transformed
 to another ideal state, of lower energy density.
 Using this time self-similar formulation, we can derive
 possible end states without time iteration,
 assuming a specified initial configuration.
 By circumventing intermediate stages, this self-similar approach,
 which is assumed to be independent of self-organization
 physical arguments, appears to be developing the required
 solutions for the problems waiting to be solved.
  
Under axisymmetry, we have established two solutions for
 equatorial ejection.
 The first solution has a finite azimuthal magnetic field,
 and represents a plasma torus with closed toroidal field lines.
 Beyond the plasma torus, the magnetic field lines are open
 and the plasma is not confined. This plasma torus provides
 the principle structure for the interacting-wind model.
 The second solution has a null azimuthal field,
 and represents a bipolar planetary nebula.
 Whether these self-organized structures continue to expand
 indefinitely depends on the integration constant $H$
 of the evolution function, which could have
 an asymptotically bounded track for a stable torus,
 and an unbounded track for an ever expanding bipolar nebulae.
 Additional nebula features, such as filaments and jets,
 could be accounted for with a fully three-dimensional
 self-similar MHD model.

We in fact believe that this description of self-organized
 plasmas, using self-similar solutions, could represent a
 fundamental process in astrophysical ejection events.
 Apart from the equatorial ejection solutions of plasma torus
 and bipolar structure of planetary nebulae discussed here,
 the highly collimated polar ejection solutions could be
 relevant to extragalactic jets, should we consider these
 objects to be ejection events (Tsui and Serbeto 2007).
 The collimated polar ejection mechanism could also be relevant
 to some asymmetric supernovae, with recoil on the neutron star,
 and to quasar ejection models of active galactic nucleus
 in cosmology.

\begin{acknowledgements}

The author is deeply grateful to Prof. Akira Hasegawa for the very
 essential concept of self-organization in fluids and plasmas.

\end{acknowledgements}

\clearpage
\newpage
\begin{figure}
\resizebox{\hsize}{!}{\includegraphics{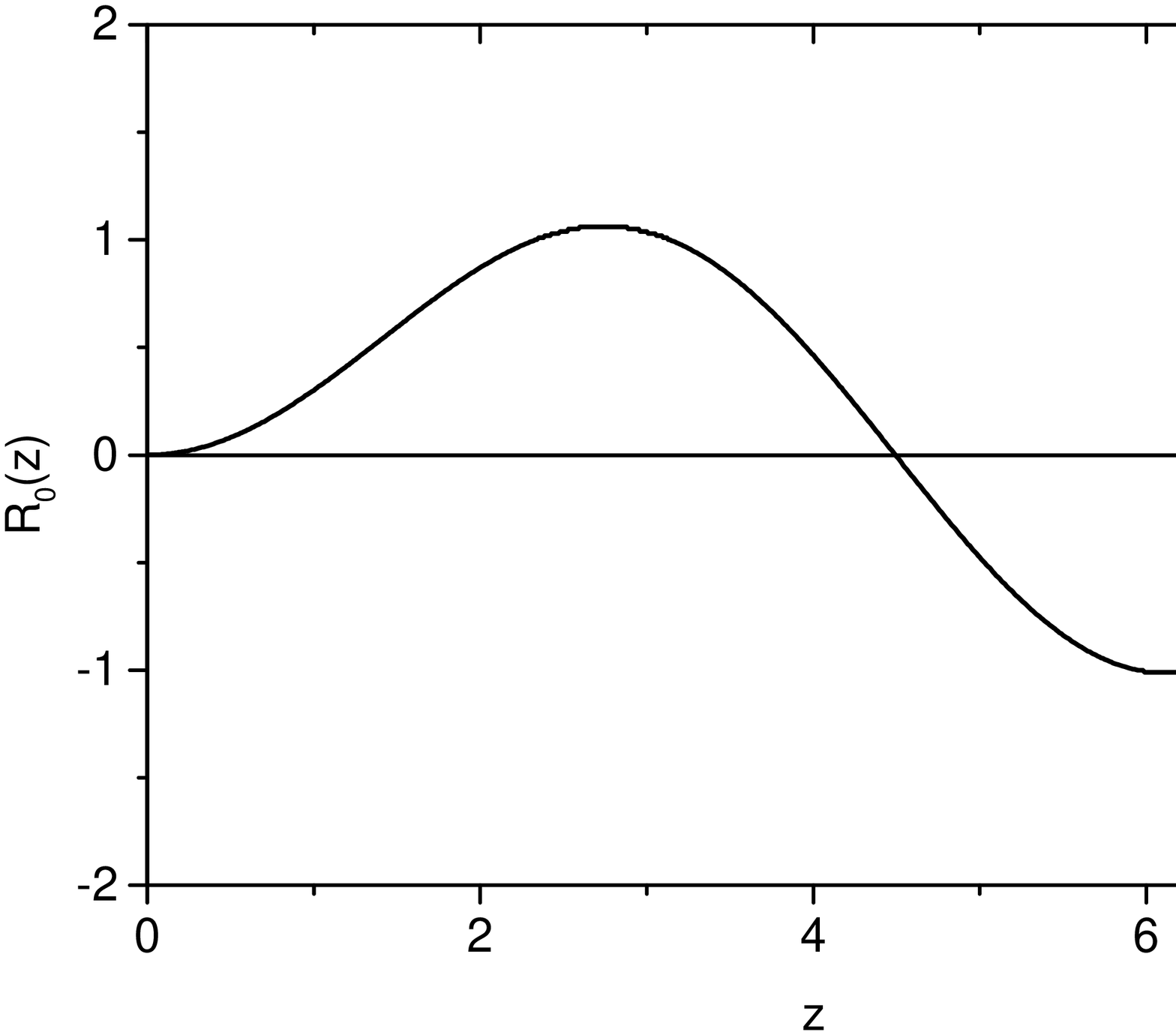}}
\caption{The function $R_{0}(z)=A_{0}zj_{1}(z)$ of the
 homogeneous solution is plotted as a function of $z=a\eta$
 to show the radial domains of the axisymmetric plasma
 structures.}
\label{fig.1}
\end{figure}

\begin{figure}
\resizebox{\hsize}{!}{\includegraphics{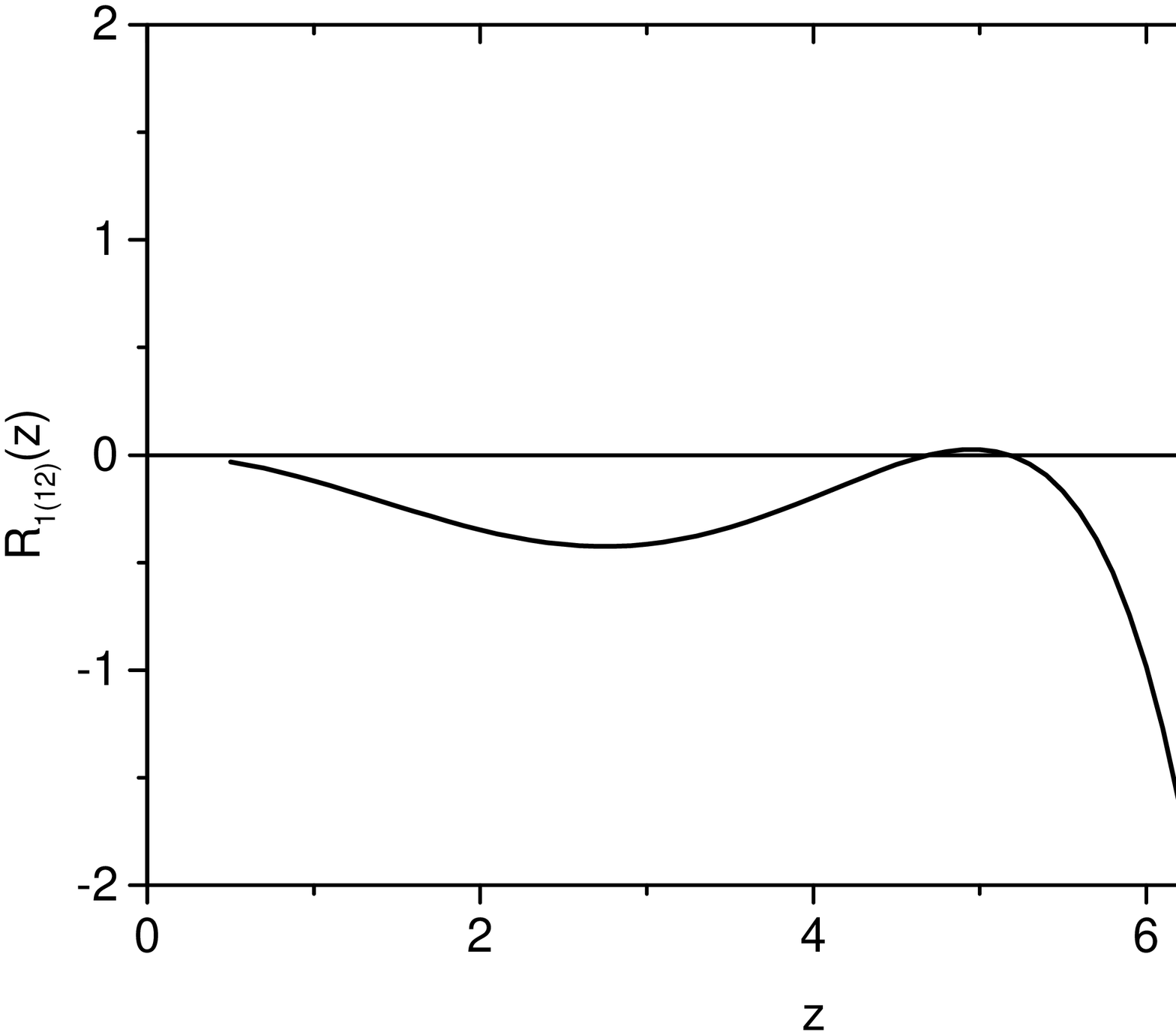}}
\caption{The power series solution $R_{1(12)}(z)$ of the
 particular solution, with $A=1\times 10^{-7}$,
 is plotted as a function of $z$
 to show the plasma pressure effect on the radial domains
 of the axisymmetric plasma structures.}
\label{fig.2}
\end{figure}

\begin{figure}
\resizebox{\hsize}{!}{\includegraphics{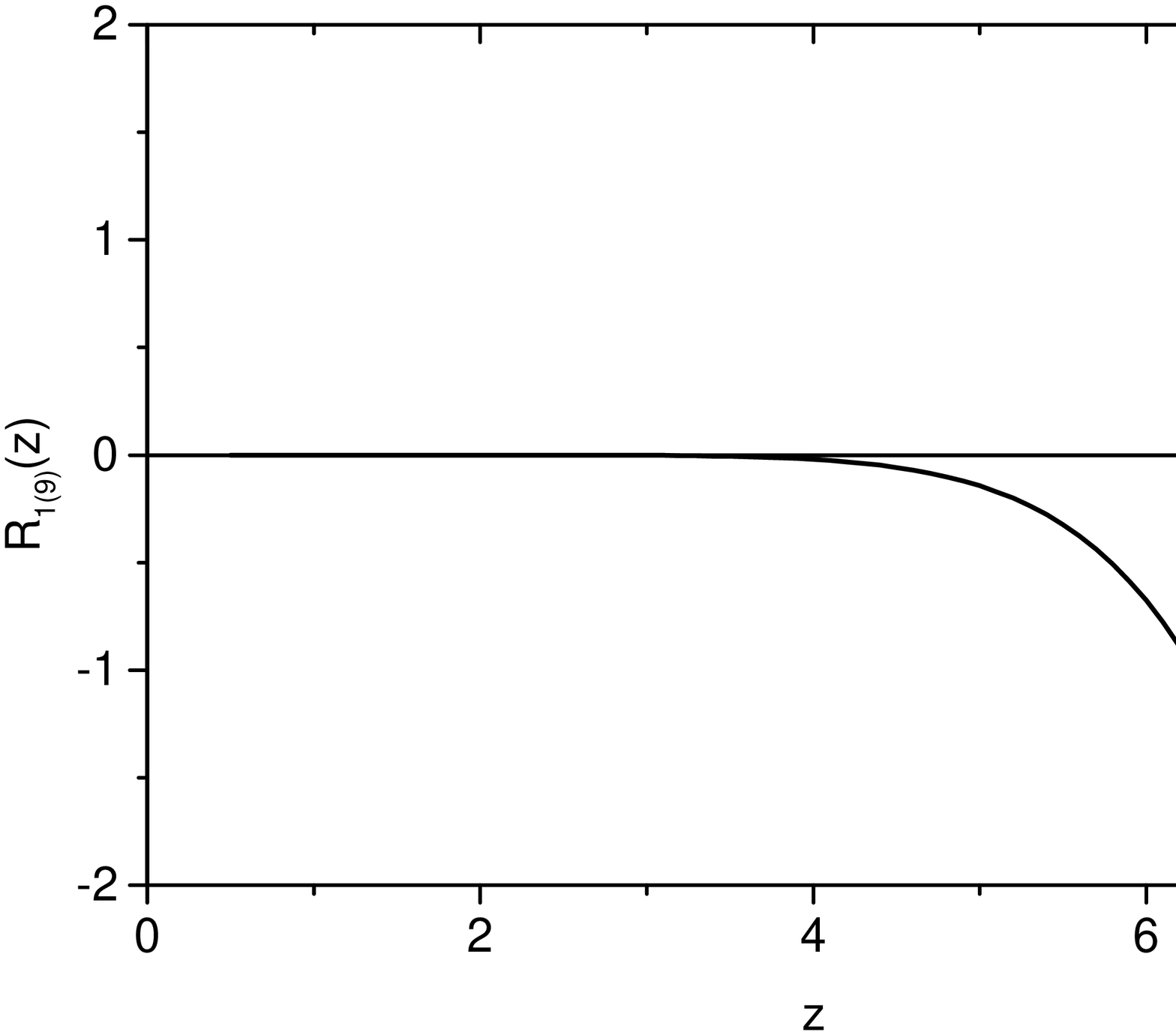}}
\caption{The power series solution $R_{1(9)}(z)$ of the
 particular solution, with $A=1\times 10^{-7}$ and $\kappa^{3}=16$,
 is plotted as a function of $z$
 to show the plasma pressure effect on the radial domains
 of the axisymmetric plasma structures.}
\label{fig.3}
\end{figure}

\begin{figure}
\resizebox{\hsize}{!}{\includegraphics{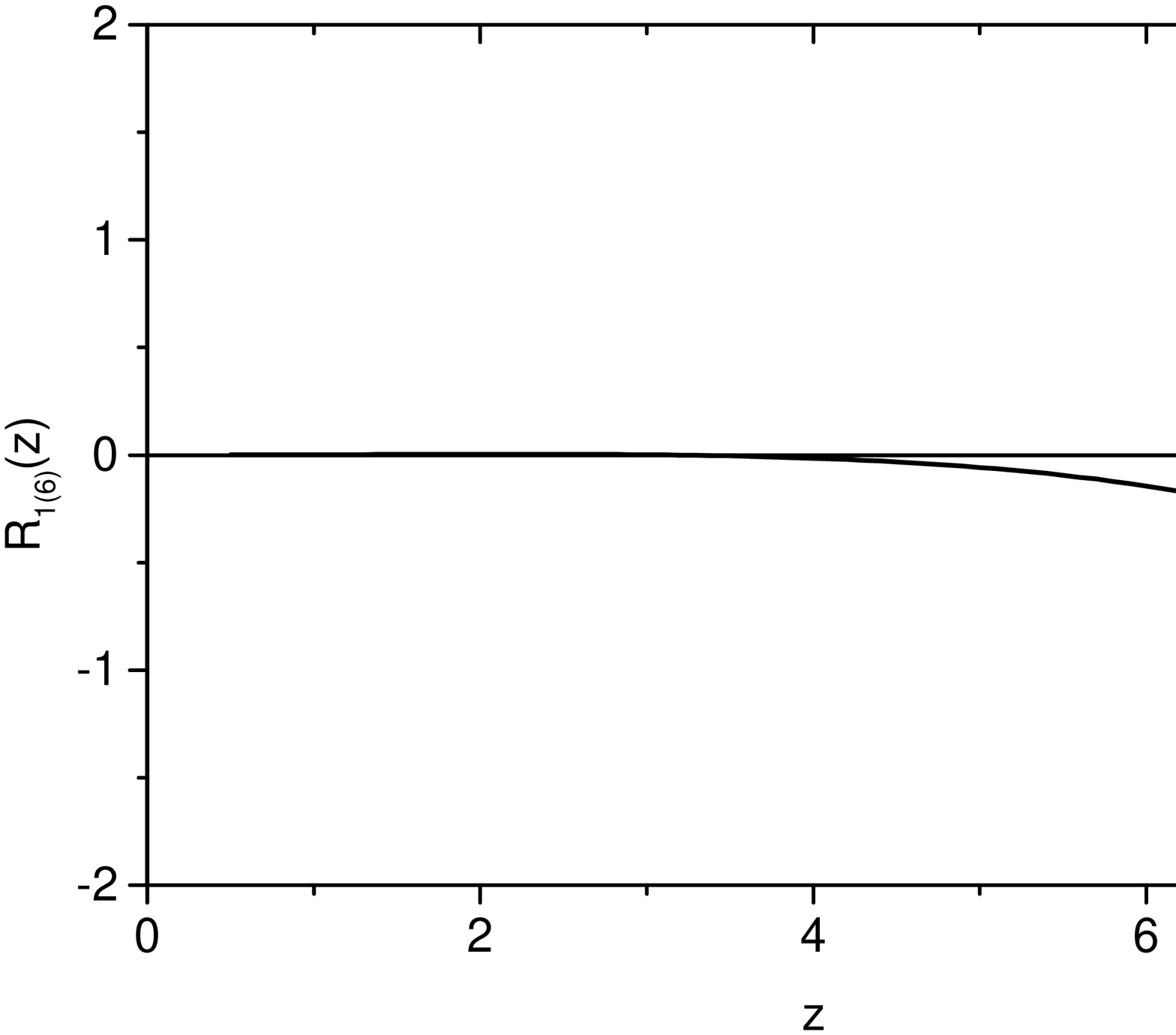}}
\caption{The power series solution $R_{1(6)}(z)$ of the
 particular solution, with $A=1\times 10^{-7}$ and $\kappa^{3}=16$,
 is plotted as a function of $z$
 to show the plasma pressure effect on the radial domains
 of the axisymmetric plasma structures.}
\label{fig.4}
\end{figure}

\begin{figure}
\resizebox{\hsize}{!}{\includegraphics{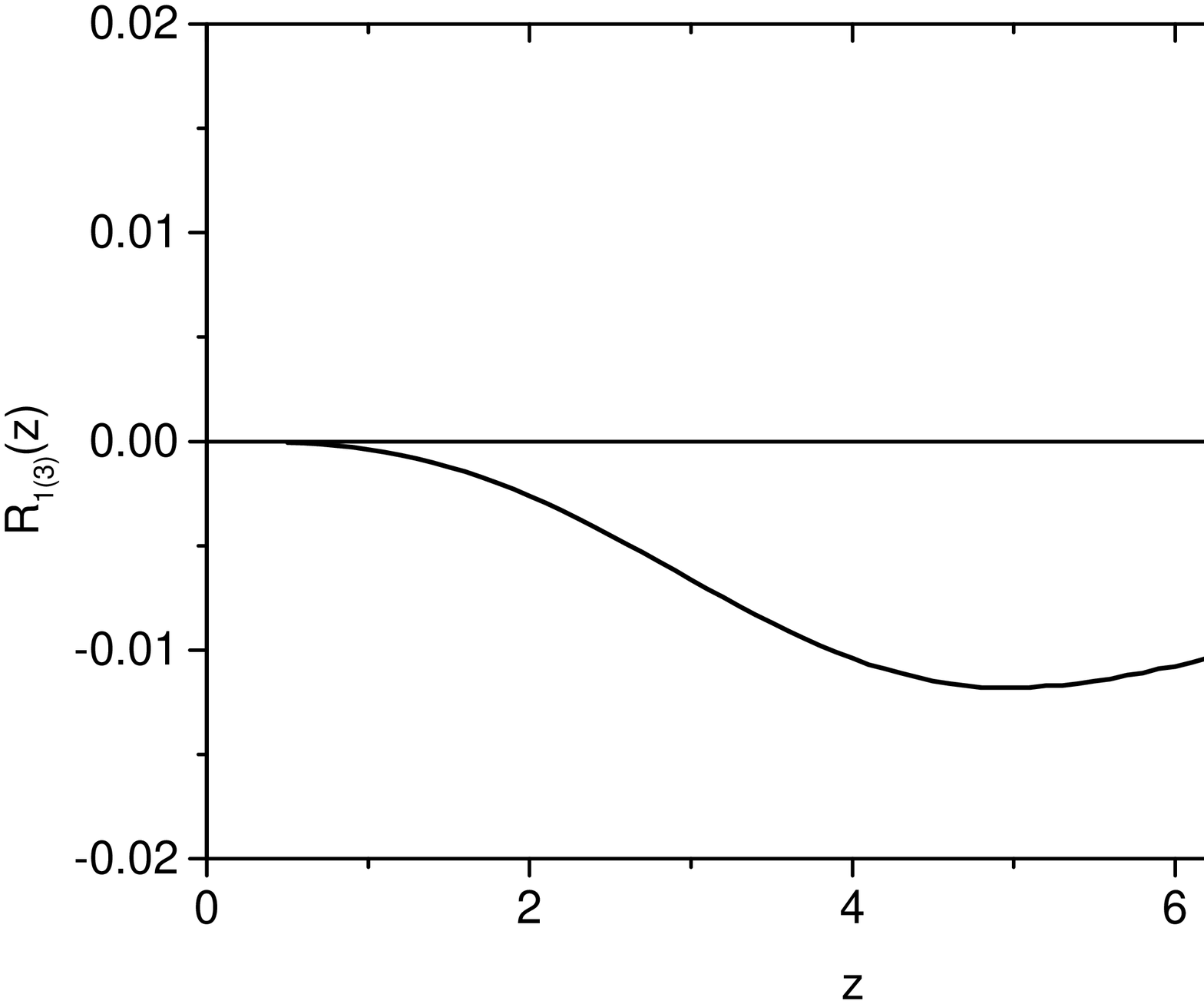}}
\caption{The power series solution $R_{1(3)}(z)$ of the
 particular solution, with $A=1\times 10^{-7}$ and $\kappa^{3}=16$,
 is plotted as a function of $z$
 to show the plasma pressure effect on the radial domains
 of the axisymmetric plasma structures.}
\label{fig.5}
\end{figure}

\begin{figure}
\resizebox{\hsize}{!}{\includegraphics{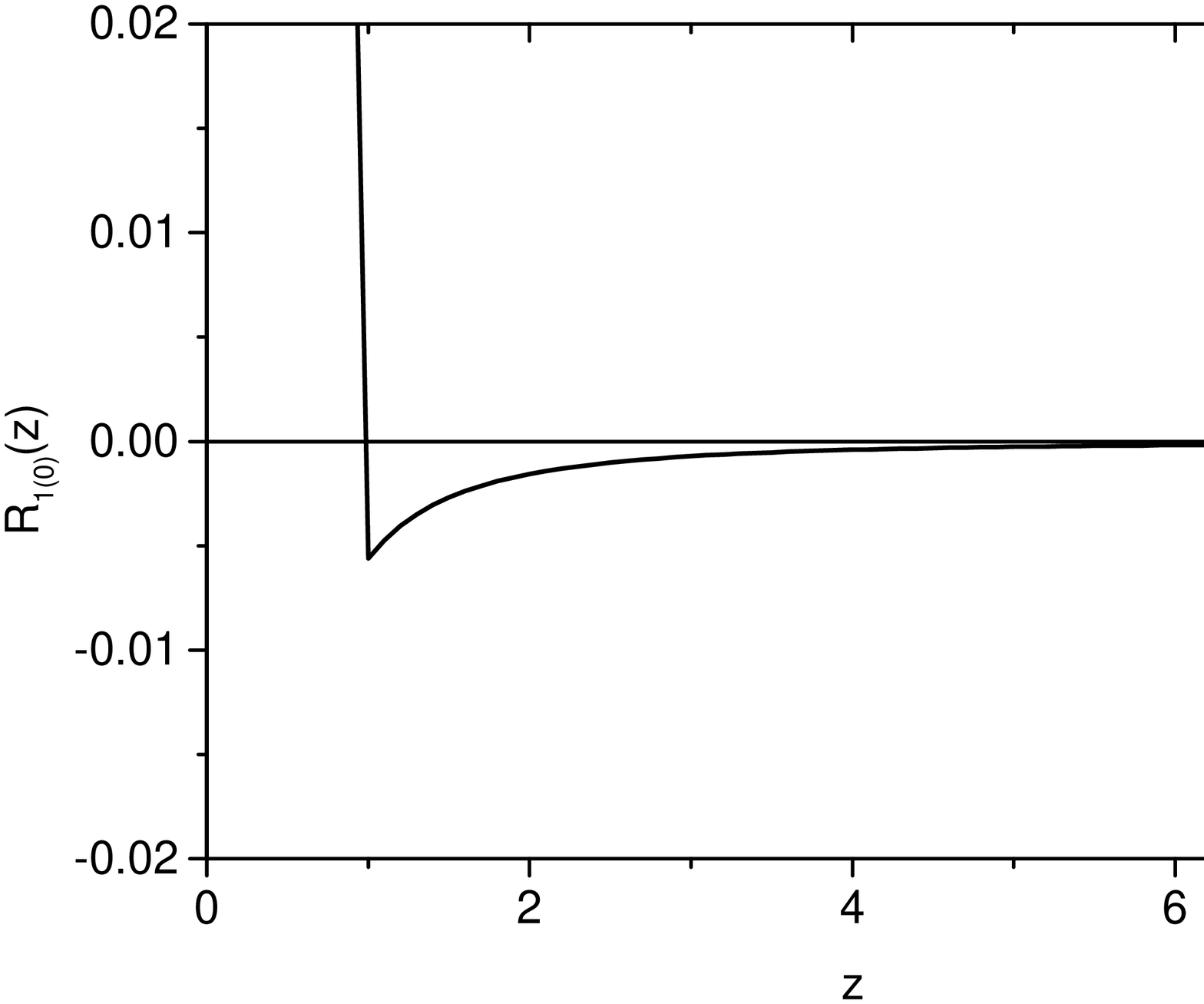}}
\caption{The power series solution $R_{1(0)}(z)$ of the
 particular solution, with $A=1\times 10^{-7}$ and $\kappa^{3}=16$,
 is plotted as a function of $z$
 to show the plasma pressure effect on the radial domains
 of the axisymmetric plasma structures.}
\label{fig.6}
\end{figure}

\begin{figure}
\resizebox{\hsize}{!}{\includegraphics{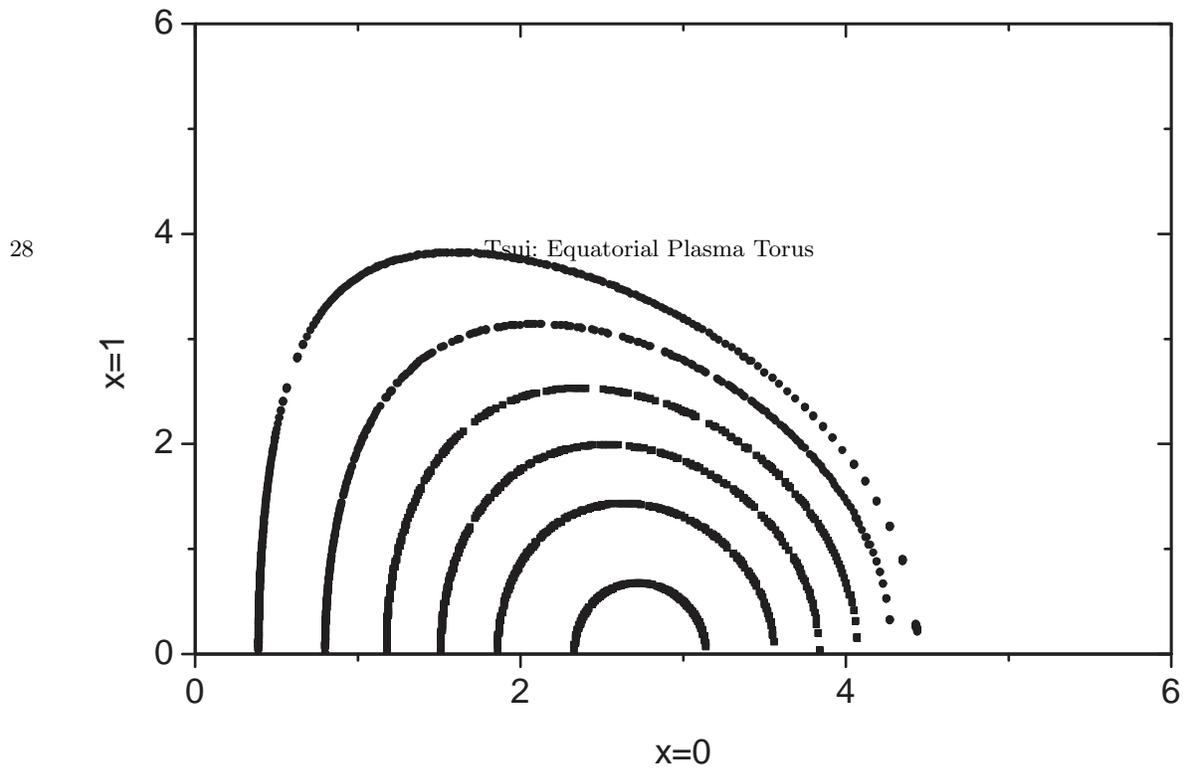}}
\caption{The magnetic field lines using the homogeneous
 solution $R_{0}(z)$ are shown with $C=+0.05$, $C=+0.20$, $C=+0.40$,
 $C=+0.60$, $C=+0.80$, and $C=+1.00$ from the outer to inner
 contours respectively.
 The axes are labelled in $x=\cos\theta$, and
 the contribution of the plasma pressure represented by $R_{1}(z)$
 is neglected here.}
\label{fig.7}
\end{figure}

\begin{figure}
\resizebox{\hsize}{!}{\includegraphics{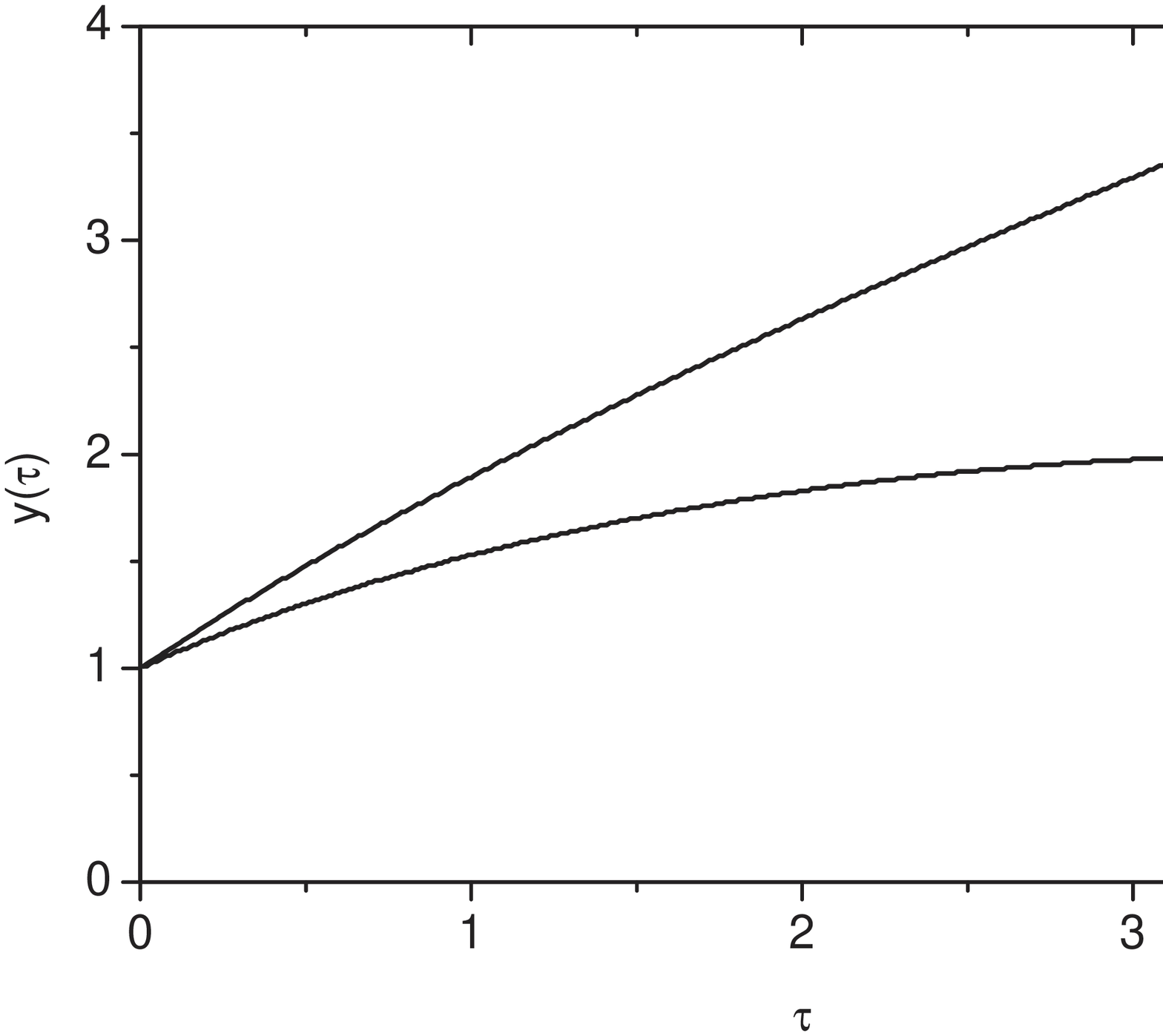}}
\caption{The asymptotically bounded evolution function,
 the lower curve, is plotted as a function of the normalized
 time $\tau$ with $|H|/\alpha=0.5$, whereas the unbounded
 evolution function, the upper curve, is plotted with
 $H/\alpha=0.1$.}
\label{fig.8}
\end{figure}

\begin{figure}
\resizebox{\hsize}{!}{\includegraphics{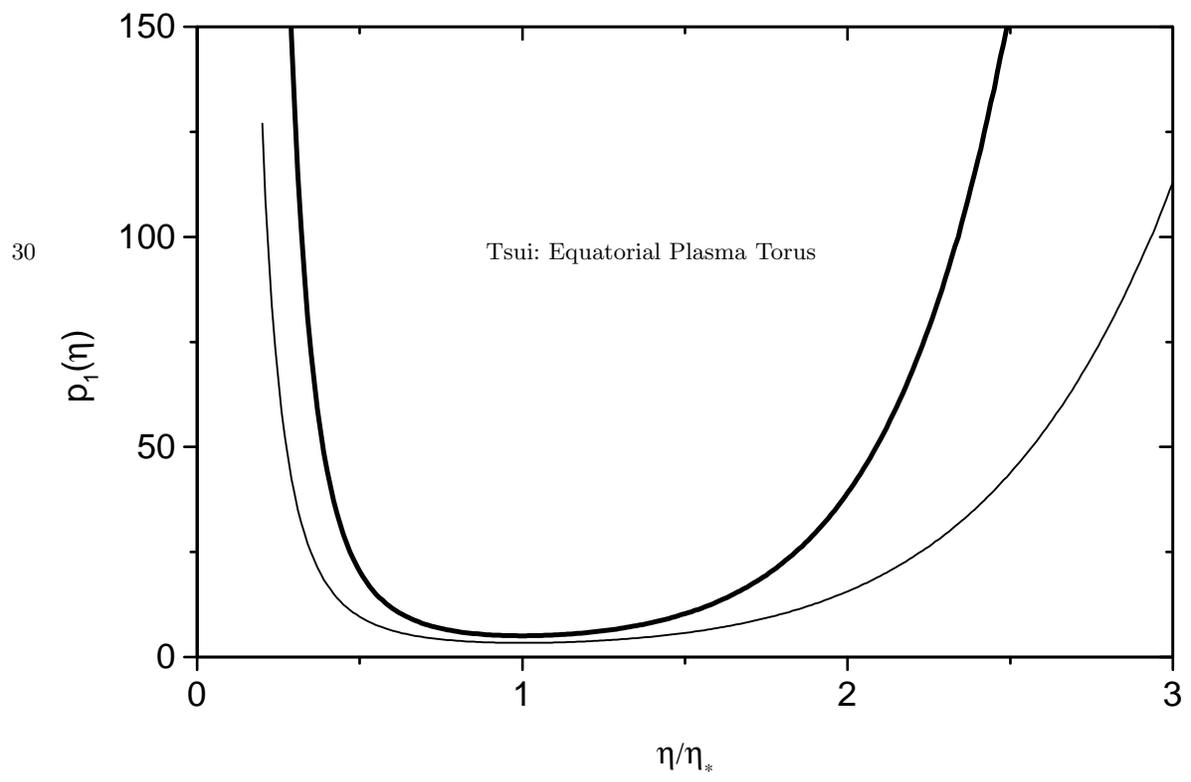}}
\caption{The explicit radial dependence of pressure profile
 $\bar p_{1}(\eta)$ is shown
 as a function of the normalized radial label $\eta/\eta_{*}$
 through the thick line.
 The explicit radial dependence of mass density profile
 $\bar\rho_{1}(\eta)$ is also shown
 through the thin line for comparisons.}
\label{fig.9}
\end{figure}

\begin{figure}
\resizebox{\hsize}{!}{\includegraphics{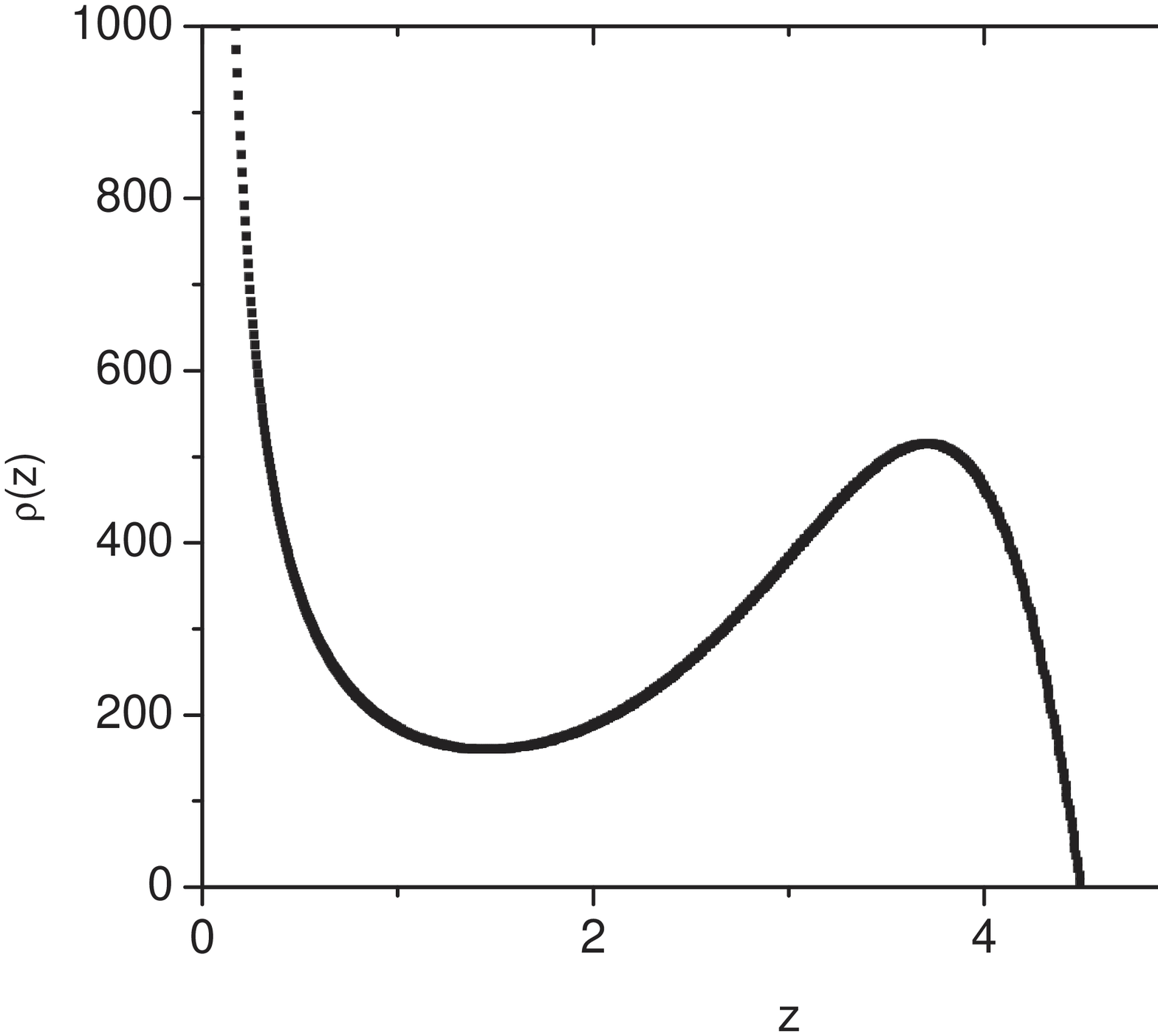}}
\caption{The full, explicit and implicit, radial mass density
 profile $\bar\rho(\eta,\bar P)$ using the homogeneous solution
 $R_{0}(z)$ is shown as a function of $z$.}
\label{fig.10}
\end{figure}

\begin{figure}
\resizebox{\hsize}{!}{\includegraphics{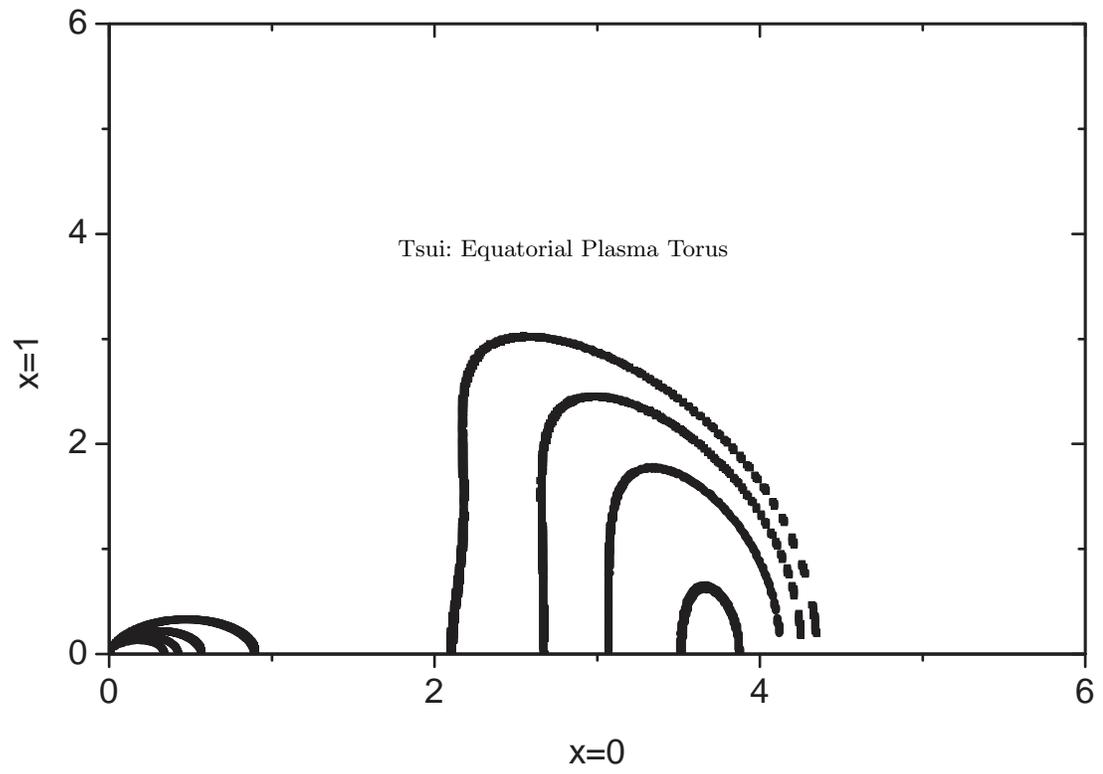}}
\caption{The mass density contours $\bar\rho(\eta,\bar P)$
 with $\bar C=0$ are shown with
 $C=+200$, $C=+300$, $C=+400$, and $C=+500$
 from the outer to inner contours respectively.}
\label{fig.11}
\end{figure}

\begin{figure}
\resizebox{\hsize}{!}{\includegraphics{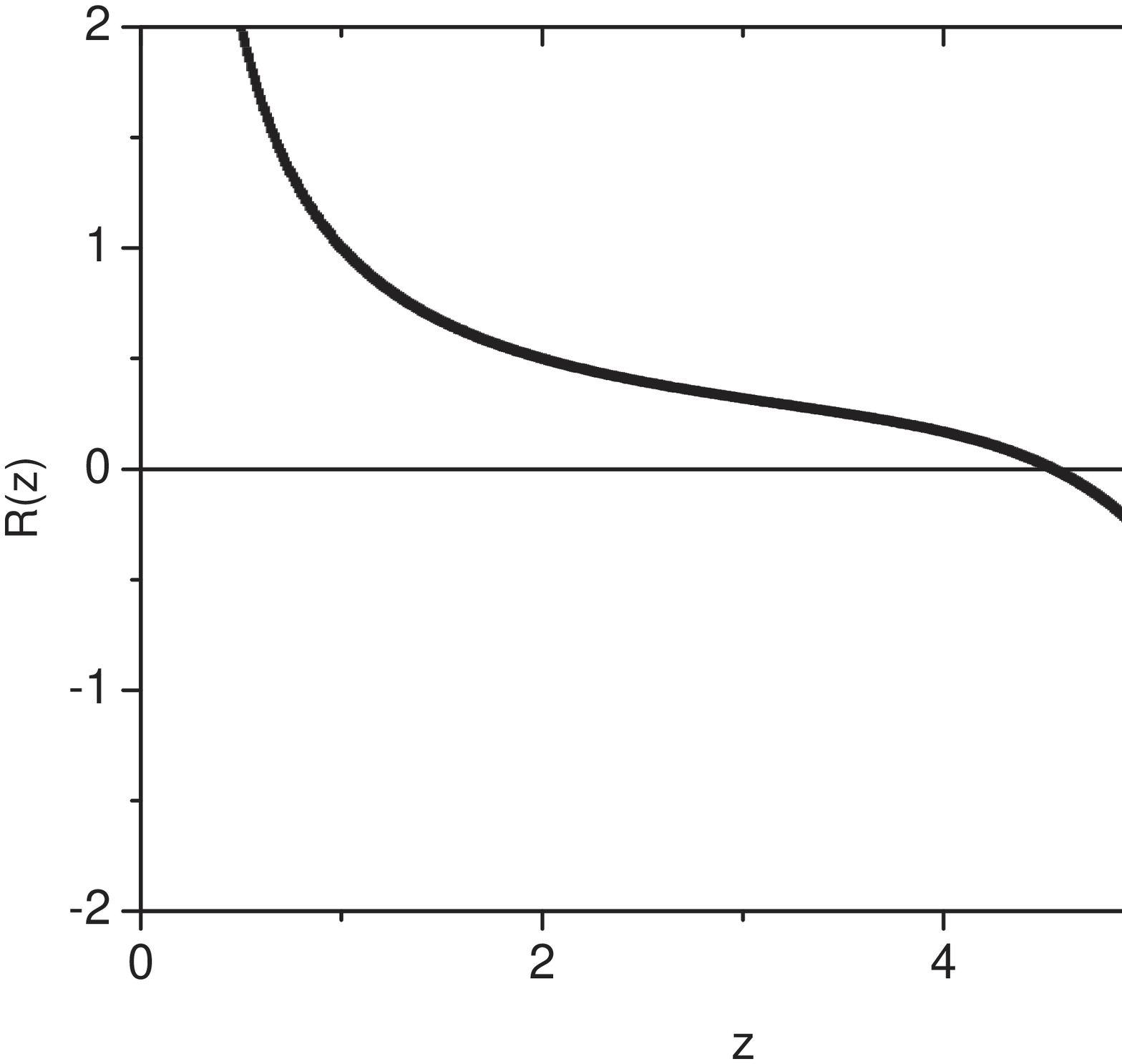}}
\caption{The general solution of the radial function $R(z)$
 is shown as a function of $z$.}
\label{fig.12}
\end{figure}

\begin{figure}
\resizebox{\hsize}{!}{\includegraphics{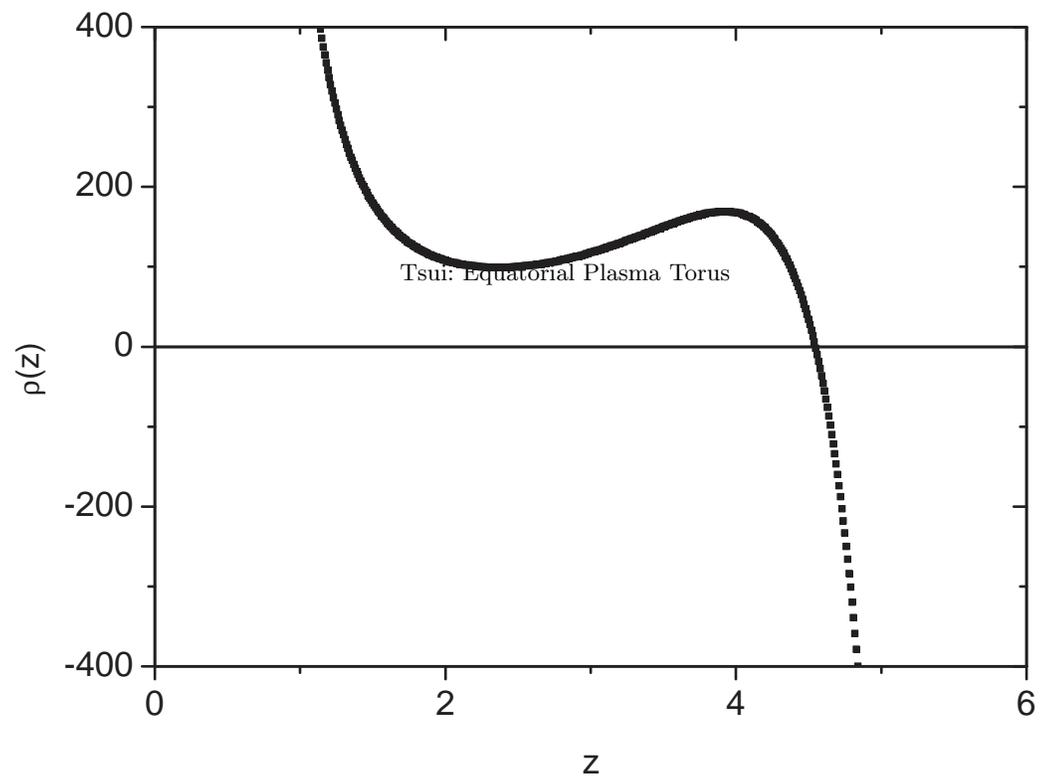}}
\caption{The full, explicit and implicit, radial mass density
 profile $\bar\rho(\eta,\bar P)$ is shown as a function of $z$.}
\label{fig.13}
\end{figure}

\begin{figure}
\resizebox{\hsize}{!}{\includegraphics{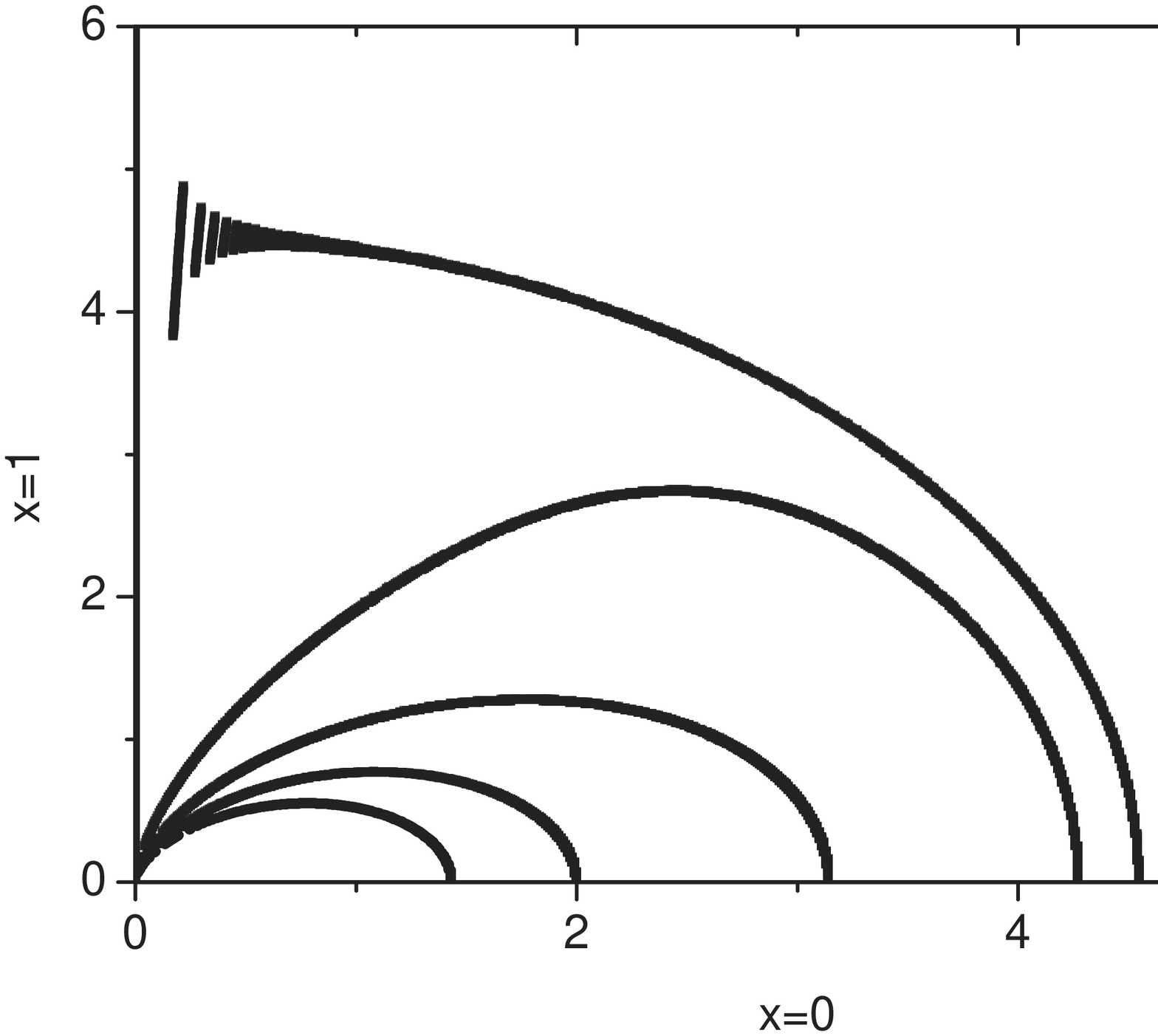}}
\caption{The magnetic field lines are shown with
 $C=+0.0$, $C=+0.1$, $C=+0.3$, $C=+0.5$, and $C=+0.7$
 from the outer to inner contours respectively.}
\label{fig.14}
\end{figure}

\end{document}